\documentclass[preprint,showpacs,preprintnumbers,amsmath,amssymb]{revtex4}
\usepackage{graphicx}% Include figure files
 \usepackage{floatflt}
 \usepackage{dcolumn}% Align table columns on decimal point
 \usepackage{bm}% bold math

\begin{document}

\preprint{APS/123-QED}
\title{ Effects of  $\omega$  meson self-coupling on the properties
of finite nuclei and neutron stars}

\author{Raj Kumar$^1$, B. K. Agrawal$^2$ and Shashi K. Dhiman$^{1}$} 
\affiliation{
$^1$Department of Physics,
H.P. University, Shimla - 171005,
India.\\
$^2$Saha Institute of Nuclear Physics, Kolkata - 700064, India.}
\begin{abstract}

The effects of $\omega$ meson self-coupling (OMSC) on the properties of
finite nuclei and neutron stars are investigated within the framework
of effective field theory based relativistic mean-field (ERMF) model
which includes the contributions from all possible mixed interactions
between the scalar-isoscalar ($\sigma$), vector-isoscalar ($\omega$) and
vector-isovector ($\rho$) mesons upto the quartic order.  For a realistic
investigation, several parameter sets corresponding to different values
of OMSC are generated by adjusting the remaining parameters of the ERMF
model to fit the properties of the finite nuclei. Though, all these
parameter sets give equally good fit to the properties of the finite
nuclei, only  moderate values of OMSC are favored from the ``naturalness"
point of view.  The equation of state  for the symmetric nuclear and
pure neutron matters resulting from the parameter sets with the moderate
values of OMSC are in close agreement with the ones obtained within the
Dirac-Brueckner-Hartree-Fock approximation. For such parameter sets the
limiting mass for the neutron stars composed of $\beta$-stable matter
is $\sim 1.9M_\odot$.  It is found that the direct Urca process can
occur in the neutron stars with ``canonical" mass of $1.4{\it M}_\odot$
only for the moderate and higher values of OMSC.  Some other interesting
properties for the neutron stars are also discussed.

\end{abstract}
\pacs{21.10.-k,21.65+f,24.30.Cz,21.60jz,26.60.+c} \maketitle

\newpage
\section{Introduction}
\label{sec:intro}
The concepts of effective field theory (EFT) have provided
a modern perspective to the  relativistic mean-field models
\cite{Furnstahl96,Serot97,Furnstahl97}.  The EFT based relativistic
mean-field (ERMF) models are obtained by expanding the energy density
functional in powers of the fields for scalar-isoscalar ($\sigma$),
vector-isoscalar ($\omega$) and vector-isovector ($\rho$) mesons and
their derivatives upto a given order $\nu$ .  Thus, a ERMF model includes
the contributions from all possible  self and mixed interaction terms
for $\sigma$, $\omega$ and $\rho$ mesons in addition to the cubic and
quartic self interaction terms for $\sigma$ meson as present in the
conventional quantum hydrodynamic based relativistic mean field   models
\cite{Walecka74,Boguta77}.  The parameters (or the expansion coefficients)
appearing in the energy density functional are	so  adjusted that the
ERMF results for a set of  nuclear observables agree well with the
corresponding experimental data.  The EFT demands that the adjusted
parameters must exhibit ``naturalness",  i.e.,	the values of all the
parameters should be  roughly of same size when expressed in appropriate
dimensionless  ratios \cite{Muller96}. The lack of ``naturalness" implies
that the omitted terms are important.  Sometimes not all the terms upto
a given order $\nu$ are considered which might lead to	 unnaturalness
\cite{Furnstahl96}.  It is found in Ref. \cite{Furnstahl97} that the
ERMF models containing terms upto order $\nu = 4$   can be satisfactorily
applied to study the properties of finite  nuclei. The inclusion of next
higher order terms improves the fit to the finite nuclear properties only
marginally.  In Ref.  \cite{Muller96} it has been shown that even high
density behaviour of the equation of state (EOS) of pure neutron matter
and $\beta$-stable matter are predominantly controlled by the quartic
order $\omega$ meson self-coupling (OMSC). The effects of inclusion of
higher order terms on the high density behaviour of EOS are found to be
only modest to negligible.

There have been  several parameterizations \cite{Serot97,Estal01}
of ERMF models containing most of the terms upto the quartic order
($\nu =4$). These parameter sets are obtained by a fit to the set of
experimental data for the properties of a few stable closed shell
nuclei. The fitted parameters exhibit ``naturalness". But, some
of the nuclear matter properties they yield requires attention;
the linear density dependence of the symmetry energy coefficient
is strong and the nuclear matter incompressibility coefficient
$K$ is either little too low or quite  high.  The  linear density
dependence of the symmetry energy coefficient and the nuclear matter
incompressibility coefficient should be adjusted to yield reasonably
the  neutron-skin thickness and the centroid energy of the isoscalar
giant monopole resonance, respectively.  The FSUGold parameter set is
obtained  recently \cite{Todd-Rutel05,Piekarewicz05} in such a way that
they give realistic values for all the properties normally associated
with the nuclear matter at the saturation density. However, some of the
parameters of FSUGold set show deviation from ``naturalness". This may be
pointing to the fact that the contributions due to the mixed  interaction
terms involving $\sigma-\omega$ and $\sigma-\rho$ mesons omitted from
the FSUGold parametrization are important.  For  FSUGold parameters,
the limiting mass of the neutron star is only 1.7${\it M}_\odot$ which
might reduce further   with the inclusion of hyperons degree of freedom
\cite{Glendenning91,Ban04}.  On the other hand, an early  study using
ERMF model shows  Ref.\cite{Muller96}  that by changing the value of
OMSC within a reasonable range, the limiting mass of neutron stars
can be varied  from   1.8${\it M}_\odot$ - 2.8$ {\it M}_\odot$ for a
fixed set of nuclear matter properties.  Of course, for a realistic
variation in the value of limiting mass, the nuclear matter properties
must be determined from the experimental data on the finite nuclei.
It is also to be  noted that in Ref.\cite{Muller96} the effects of the
mixed interactions were not considered and no constrain was imposed on
the density dependence of symmetry energy coefficient.

In the present work we have performed a realistic investigation involving
the effects of OMSC on the properties of finite nuclei and neutron
stars using a ERMF model. We consider a  ERMF model which includes
the contributions from all  possible mixed interactions between  the
$\sigma$, $\omega$ and $\rho$ mesons upto the quartic order. Like, in
the case of FSUGold parametrization, we determine all the properties
normally associated with nuclear matter  from the experimental data on
the finite nuclei. We generate  several parameter sets for fixed values
of OMSC by adjusting the remaining parameters of the ERMF model to fit
exactly the same set of experimental data for the total binding energies
and charge rms radii for some closed shell nuclei.  The binding energy
data is considered for nuclei ranging from normal to the exotic  ones.
We also  include in our fit the value of the neutron-skin thickness for
$^{208}$Pb nucleus to constrain the linear density dependence of the
symmetry energy coefficient. We restrict the value of the nuclear matter
incompressibility coefficient to be within $220 -240$ MeV as required
by the experimental data on the centroid energies for the isoscalar
giant monopole resonance.  The best fit parameters are searched using
the simulated annealing method (SAM) which we have applied recently
\cite{Agrawal05,Agrawal06} to determine the parameters of the standard
and generalized Skyrme type effective forces.

In Sec. \ref{sec:eft} we discuss ERMF model in brief. In
Sec. \ref{sec:sam} we discuss the simulated annealing method used  in the
present work to search for the best fit parameters. Several parameter
sets for the different values of the OMSC generated in the present
work are given in Sec. \ref{sec:new_par}. In Secs. \ref{sec:snm_pnm} -
\ref{sec:ns} we discuss our results for the symmetric nuclear matter,
pure neutron matter, finite nuclei and neutron stars obtained using
our newly generated parameter sets and compare them with the ones
for  the FSUGold parametrization. As a customary, we also compare our
results with those obtained using most popularly used NL3 parameter set
\cite{Lalazissis97}. Finally, in Sec.  \ref{sec:conc} we summarize our
main conclusions.

\section{Energy density functional}
\label{sec:eft}
The derivations of the EFT motivated effective Lagrangian and
corresponding energy density functionals  are well documented in Refs.
\cite{Furnstahl96,Serot97,Furnstahl97}. In this section we shall outline
very  briefly the final expressions for the energy density functional
and field equations within the mean-field approximations as used in our
numerical calculations. The energy density functional containing all the
self and mixed interaction terms for the $\sigma$, $\omega$ and $\rho$
mesons upto the quartic order reads as \cite{Muller96,Serot97},
\begin{equation}
\label{eq:eden}
\begin{split}
{\cal E}(r)&=\sum_{\mu}\phi_\mu^\dagger\left \{
-i {\bf \alpha}\cdot{\bf \nabla} +\beta\left[M -\Phi (r)\right]
+ W(r)+ \frac{1}{2}\tau_3 R(r)+ \frac{1+\tau_3}{2}A(r)\right\}\phi_\mu
(r)\\ 
&+\frac{1}{2}\frac{m_{\sigma}^2}{g_{\sigma}^2}\Phi^2(r)+
\frac{\overline{\kappa}}{6}\Phi^3(r)+ \frac{\overline{\lambda}}{24}\Phi^4(r)
-\frac{\zeta}{24}W^4(r)- \frac{\xi}{24}R^4(r)+\frac{1}{2 g_{\sigma}^2}{\left [\nabla
\Phi(r)\right]}^2\\
& -\frac{1}{2 g_{\omega}^2}{\left [\nabla W(r)\right]}^2
-\frac{1}{2}\frac{m_{\omega}^2}{g_{\omega}^2}W^2(r)-\overline{\alpha}_{1}
\Phi(r) W^2(r) -\frac{1}{2}\overline{\alpha}_{1}^\prime \Phi^2(r) W^2(r)
-\frac{1}{2}\frac{m_{\rho}^2}{g_{\rho}^2}R^2(r)\\
&-\frac{1}{2 g_{\rho}^2}
{\left[\nabla R(r)\right]}^2 
+\overline{\alpha}_{2}\Phi(r)R^2(r)-\frac{1}{2}
\overline{\alpha}_{2}^\prime
\Phi^2(r)R^2(r)-\frac{1}{2}\overline{\alpha}_{3}^\prime W^2(r)R^2(r)\\
& -\frac{1}{2
e^2}{\left[\nabla A(r)\right]}^2    
\end{split}
\end{equation}
where, the index $\mu$ runs over all occupied states of the positive
energy spectrum. The variables $\Phi$, $W$and $R$  represent the $\sigma$,
$\omega$ and $\rho$ meson fields, respectively,  and $A$ is the photon
field.  We must emphasize at this point that in the present work we
consider all the mixed interaction terms.

The mean-field equations for the nucleons, mesons and photons are as follows,

\begin{equation}
\label{eq:nucleon}
\left \{-i {\bf \alpha} \cdot {\bf \nabla} + \beta \left [ M - \Phi(r)\right ] +
W(r)+\frac{1}{2}\tau_3R(r)+\frac{1+\tau_3}{2}A(r)\right\}\phi_\mu(r)=\epsilon_\mu\phi_\mu(r),
\end{equation}

\begin{equation}
\label{eq:phi}
\begin{split}
-\Delta \Phi(r) + m_\sigma^2\Phi(r)&=
g_\sigma^2\rho_s(r)-\frac{\overline{\kappa}}{2}\Phi^2(r)-\frac{\overline{\lambda}}{6}\Phi^3(r)+\overline{\alpha}_1W^2(r)\\
&+\overline{\alpha}_1^\prime\Phi(r)
W^2(r)
+\overline{\alpha}_2R^2(r)+\overline{\alpha}_2^\prime \Phi(r) R^2(r),
\end{split}
\end{equation}

\begin{equation}
\label{eq:w}
\begin{split}
-\Delta W(r) + m_\omega^2W(r)&=
g_\omega^2\rho(r)-\frac{\zeta}{6}W^3(r)-2
\overline{\alpha}_1\Phi(r)W(r)\\
&-\overline{\alpha}_1^\prime\Phi^2(r)
W(r)
-\overline{\alpha}_3^\prime W(r)R^2(r),
\end{split}
\end{equation}

\begin{equation}
\label{eq:r}
\begin{split}
-\Delta R(r) + m_\rho^2R(r)&=
\frac{1}{2}g_\rho^2\rho_3(r)-\frac{\xi}{6}R^3(r)-2
\overline{\alpha}_2\Phi(r)R(r)\\
&-\overline{\alpha}_2^\prime\Phi^2(r)
R(r)-\overline{\alpha}_3^\prime R(r)W^2(r),
\end{split}
\end{equation}

\begin{equation}
\label{eq:photon}
-\Delta A(r) =
e^2\rho_p(r).
\end{equation}

The different kinds of baryon densities $\rho_s$ (scalar), $\rho$ (total),
$\rho_3$ (isovector) and $\rho_p$ (proton) appearing in Eqs. (\ref{eq:phi}) -
(\ref{eq:photon}) are given by,
\begin{eqnarray}
\rho_s(r) = \sum_{\mu}\phi_{\mu}^\dagger(r)\beta\phi_{\mu}(r),\\
\rho(r) = \sum_{\mu}\phi_{\mu}^\dagger(r)\phi_{\mu}(r),\\
\rho_3(r) = \sum_{\mu}\phi_{\mu}^\dagger(r)\tau_3\phi_{\mu}(r),\\
\rho_p(r) = \sum_{\mu}\phi_{\mu}^\dagger(r)\left(\frac{1+\tau_3}{2}\right)\phi_{\mu}(r).
\end{eqnarray}
The EFT imposes the condition of ``naturalness" on the parameters
or the expansion coefficients  appearing in Eq. (\ref{eq:eden}). The
``naturalness" implies that the coefficients of the various terms in the
energy density functional, when expressed in appropriate dimensionless
ratios, should all be of the same size. The dimensionless ratios are
obtained by dividing the Eq. (\ref{eq:eden}) by $M^4$ and expressing
each  term in powers of  $\Phi/M$, $W/M$ and $2R/M$ \cite{Muller96}.
This means that the dimensionless ratios , namely,

\begin{equation}
\label{eq:scl_par}
\begin{split}
&\frac{1}{2c_{\sigma}^2M^2},\>
   \frac{1}{2c_{\omega}^2M^2},\>   \frac{1}{8c_{\rho}^2M^2},\>
 \frac{\overline{\kappa}}{6M}, \>   \frac{\overline{\lambda}}{24},\>\\
& \frac{\zeta}{24},
 \frac{\xi}{384},\> \frac{\overline{\alpha}_1}{M},\>
 \frac{\overline{\alpha}_1^\prime}{2},\>
\frac{\overline{\alpha}_2}{4M},\>
  \frac{\overline{\alpha}_2^\prime}{8},\>   
\frac{\overline{\alpha}_3^\prime}{8}
\end{split}
\end{equation}
should be roughly of same size.  In the above equation $c_i^2 =
g_i^2/m_i^2$, where $i$ represents $\sigma$, $\omega$ and  $\rho$ mesons.

\section{SAM algorithm for $\chi^2$ minimization}
\label{sec:sam}

The  SAM is analogous to an  annealing process in which a metal,
initially at high temperature and disordered, slowly cools so that
the system at any time is in a  thermodynamic equilibrium. As cooling
proceeds, the system becomes more ordered and approaches a frozen
ground state at zero temperature.  It is an elegant technique  for
optimization problems of large scale, in particular, where a desired
global extremum is hidden among many local extrema.  The SAM has
been found to be an extremely useful tool for a wide variety of
minimization problems of large non-linear systems in many different
areas of science (e.g., see Refs. \cite{Patrik84,Ingber89,Cohen94}).
Recently \cite{Burvenich02,Burvenich04}, the SAM was used to generate
some initial trial parameter sets for the  point coupling variant of
the relativistic mean-field model.

In the present work we use SAM to minimize the $\chi^2$ function defined
as,
 \begin{equation} \chi^2 =  \frac{1}{N_d - N_p}\sum_{i=1}^{N_d}
\left (\frac{ M_i^{exp} - M_i^{th}}{\sigma_i}\right )^2 \label {eq:chi2}
\end{equation}
where, $N_d$ is the number of  experimental data points and $N_p$ the
number of fitted  parameters. The $\sigma_i$ stands for theoretical
error and $M_i^{exp}$ and $M_i^{th}$ are the experimental and the
corresponding theoretical values, respectively, for a given observable.
Since, the $M_i^{th}$ in Eq.(\ref{eq:chi2})  is calculated using the ERMF
model, the values of $\chi^2$ depends on the values of the  parameters
appearing in Eq. (\ref{eq:eden}).  In Ref.\cite{Agrawal05} we have
described in stepwise manner the implementation of the SAM algorithm to
determine the best fit parameters for the standard Skyrme type effective
forces.  To apply SAM one needs to specify the range for each of the
parameters needs to be fitted and a set of guess parameters. In Ref.
\cite{Agrawal05} the range of the Skyrme parameters were specified
in terms of the properties of the  nuclear matter at the saturation
density. This was possible, because, the number of the Skyrme parameters
to be fitted were directly related  to the equal number of nuclear matter
properties which led to a drastic reduction in the parameter space so that
the best fit parameters could be searched efficiently. For the  case of
generalized Skyrme type effective forces we used the SAM algorithm by
specifying the parameter space directly in terms of the range of each
of the parameters \cite{Agrawal06}. Since, the number of parameters for
the generalized Skyrme type effective forces are larger than the number
of the quantities normally associated with the nuclear matter.  For the
ERMF model we have used  rather a hybrid approach.  
The parameters $g_\sigma$, $g_\omega$, $\overline{\kappa}$ and
$\overline{\lambda}$ of Eq.(\ref{eq:eden}) for a given  values
of  $\overline{\alpha}_1$, $\overline{\alpha}_1^\prime$ and
$\zeta$  are expressed in terms of the binding energy per nucleon
$\epsilon$, incompressibility coefficient $K$, effective mass  $M^*$
and saturation density $\rho_0$ for the nuclear matter as follows
\cite{Bodmer91,Furnstahl96},

\begin{equation}
C_\sigma^2 =\frac{ \frac{1}{2}\Phi^2}
{\frac{1}{2}\Phi^2U^{\prime\prime}-3\Phi U^\prime + 6U}
\label{eq:cs2}
\end{equation}

\begin{equation}
C_\omega^2 = \frac{W}{\rho - W(2\overline\alpha_1 \Phi+\overline\alpha_1^\prime \Phi^2) - \frac{1}{6}\zeta W^3}
\label{eq:cw2}
\end{equation}
\begin{equation}
\overline\kappa = \frac{-\Phi^2U^{\prime\prime}+5\Phi U^{\prime}-8U}
{\frac{1}{6}\Phi^3}
\label{eq:kappa}
\end{equation}
\begin{equation}
\overline\lambda=\frac{\frac{1}{2}\Phi^2U^{\prime\prime}-2\Phi U^\prime+3U}
{\frac{1}{24}\Phi^4}
\label{eq:lambda}
\end{equation}
where,
\begin{equation}
\Phi = M - M^*
\label{eq:phi1}
\end{equation}
\begin{equation}
W = \epsilon + M - \sqrt{k_{F}^2 + M^{*2}}
\label{eq:w1}
\end{equation}
\begin{equation}
U^\prime = \rho_{s} + W^2\left [\overline\alpha_1 +\overline\alpha_1^\prime\Phi\right ]
\label{eq:u1}
\end{equation}
\begin{equation}
U^{\prime\prime} = \rho_{s}^\prime -l^2v + \overline\alpha_1^\prime W^2 + \frac{\left (\frac{M^*}{E_{F}^*}-lv\right )^2}
{\frac{\pi^2}{2k_{F}E_{F}^*}+v-\frac{K}{9\rho}}
\label{eq:u2}
\end{equation}
with
\begin{equation}
l= -2W\left ( \overline\alpha_1 + \overline\alpha_1^\prime\Phi\right )
\label{eq:l}
\end{equation}
\begin{equation}
v=\frac{1}{\frac{1}{C_\omega^2} + \left ( 2\overline\alpha_1 +
\overline\alpha_1^\prime\Phi^2\right )+\frac{1}{2}\zeta W^2}
\label{eq:v}
\end{equation}
In Eqs. (\ref{eq:cs2}) - (\ref{eq:v}), all the quantities are obtained at
the nuclear matter saturation density $\rho =  \rho_0$.

In Table \ref{tab:prange} we specify
the parameter space in terms of $\epsilon$, $K$, $M^*$ , $\rho_0$
and some other parameters. It can be easily verified  using  second
and third columns of Table \ref{tab:prange} that the range of each of
the parameters $\overline{\alpha}_1$, $\overline{\alpha}_1^\prime$,
$\overline{\alpha}_2$, $\overline{\alpha}_2^\prime$ and
$\overline{\alpha}_3^\prime$ becomes identical ($0.0 - 2.0\times
10^{-3}$) when expressed in appropriate dimensionless ratios  using
Eq.(\ref{eq:scl_par}). The column labeled ``d" denotes the maximum
change allowed in a single step for a randomly selected parameter. The
guess values of the parameters which we have used for initiating the
SAM are given in last column of Table \ref{tab:prange}.  The values
of guess parameters are so chosen that they lie in the middle of the
specified range. It should be noted that the parameters $\zeta$ and $\xi$
denoting the self couplings for $\omega$ and $\rho$ mesons do not appear
in Table \ref{tab:prange} as they are kept fixed during $\chi ^{2}$
minimization procedure .  The best fit parameters for which $\chi^2$
is the minimum are searched within the specified parameter space with
the help of annealing schedule.  The annealing schedule determines the
value of the control parameter $T(k)$ at $k-$th step starting from its
initial value $T(0)$. As in Refs.  \cite{Agrawal05,Agrawal06} we use
Cauchy annealing schedule given by, \begin{equation} \label{eq:ann_sch}
T(k)=\frac{T(0)}{k+1} \end{equation}
where, starting value of $k$  is equal to zero and is increased in steps
of unity after each $100N_p$ reconfigurations or $10N_p$ successful
reconfigurations which ever occurs first.  We keep on reducing the
value of $T$ using Eq. (\ref{eq:ann_sch})   until the efforts to reduce
further the value of $\chi^2$  becomes sufficiently discouraging.
We have  performed  test calculations for two different values of T(0)
taken to be $ 2.5$ and $5.0$. For both the test calculations the final
value of the $\chi^2$ function is almost the same. In what follows,
we present our results obtained using $T(0) = 5.0$.

\section{New parametrizations of the ERMF model}
\label{sec:new_par}
We search for the best fit  parameters of the ERMF models   by
following  the $\chi^2$ minimization procedure as briefly outlined
in Sec.  \ref{sec:sam}.  Our data set contains the total binding
energies and charge  rms radii for several nuclei taken from Refs
\cite{Audi03,Otten89,Vries87}.  We consider total binding energies
for $^{16,24}$O, $^{40,48}$Ca, $^{56,78}$Ni, $^{88}$Sr, $^{90}$Zr,
$^{100,116,132}$Sn and $^{208}$Pb nuclei, charge rms radii for
$^{16}$O, $^{40,48}$Ca, $^{56}$Ni, $^{88}$Sr, $^{90}$Zr, $^{116}$Sn
and $^{208}$Pb nuclei .  The theoretical error needed to evaluate the
$\chi ^{2}$ function ( Eq.  \ref{eq:chi2}) are taken to be 1.0 MeV
for the total binding energies and 0.02 fm for the charge rms radii.
In addition, we also fit the value of  neutron-skin thickness for
$^{208}$Pb nucleus to constrain the linear density dependence of
symmetry energy coefficient.  The very accurate data of neutron-skin
thicknesses are still not available.  It is shown \cite{Piekarewicz04}
in a relativistic mean-field based random phase approximation that the
neutron-skin thickness of $0.175$ fm in $^{208}$Pb nucleus and  $K\approx
240$ MeV are required to adequately reproduce the centroid energies
of the isoscalar giant monopole and isovector giant dipole resonances.
A Dirac-Brueckner-Hartree-Fock (DBHF) calculation \cite{Alonso03}  using
a realistic potential predicts the value of neutron-skin thickness
in $^{208}$Pb nucleus to be $0.188$ fm.  However, recently extracted
value of neutron-skin thickness for $^{208}Pb$ nucleus from the isospin
diffusion data lie within $0.16 - 0.24$fm indicating large uncertainties
\cite{Chen05}.  In our fit we use the value of neutron-skin thickness to
be $0.18\pm 0.01$ fm for the $^{208}$Pb nucleus.  To this end, we must
point out that for our parametrizations, the center of mass correction
to the total binding energy $E_{\rm cm}$ is evaluated within the harmonic
oscillator approximation which gives $E_{\rm cm} = \frac{3}{4}\hbar\omega$
and we take $\hbar\omega = 45A^{-1/3} - 25A^{-2/3}$ MeV.

The  parameters $\zeta$ and $\xi$ corresponding to self-couplings
for $\omega$ and $\rho$ mesons can not be very well constrained by
the properties of finite nuclei. These parameters mainly determine
the high density behaviour of the EOS .  The impact of the parameter
$\xi$  is found to be appreciable for pure neutron matter only at very
high densities \cite{Muller96}.  The seven new parameter sets FSUGZ00,
FSUGZ01,....,FSUGZ06 are generated for fixed values of OMSC parameter
$\zeta = 0.00, 0.01, ....,0.06$ and keeping $\xi =0$. All the FSUGZ
parametrizations give equally good fit to the properties of the finite
nuclei. In Table \ref{tab:pfit} we give the values only for FSUGZ00,
FSUGZ03 and FSUGZ06 parameter sets.  In Table \ref{tab:dim_par} we
present the values of these parameters when expressed in dimensionless
ratios using Eq. (\ref{eq:scl_par}).  In Tables \ref{tab:pfit} and
\ref{tab:dim_par} we also list the values of the parameters for NL3
\cite{Lalazissis97} and FSUGold \cite{Todd-Rutel05,Piekarewicz05}sets.
It is clear from  the Table \ref{tab:dim_par} that for the case of
FSUGold parameter set the $\overline{\alpha}_3^\prime$ is quite large in
comparison to the other parameters when expressed in appropriate
dimensionless ratios.  We notice from Table \ref{tab:dim_par} that the
values of the parameters $\overline{\kappa}$, $\overline{\lambda}$,
$\overline{\alpha}_1$ and $\overline{\alpha}_1^\prime$ are strongly
correlated with the values of $\zeta$. For smaller  $\zeta$, the
parameter $\overline{\lambda}$ becomes small and negative which not only
gives rise to deviations from the ``naturalness" behaviour but also
gives instabilities in the EOS at high densities. Whereas, for larger
$\zeta$, the values of $\overline{\kappa}$, $\overline{\alpha}_1$ and
$\overline{\alpha}_1^\prime$ become smaller and once again show deviations
from the ``naturalness".  Thus, it appears that  a moderate value of
$\zeta \sim 0.03$  is favored from the ``naturalness" point of view.

To understand better, the overall  ``naturalness" behaviour of the various
parametrizations  of the ERMF model presented in Table \ref{tab:dim_par},
we consider the ratio of the largest to the smallest parameters for a
given set.  One expects this ratio to be  of the order of unity if the
parameters strictly obey  the ``naturalness". We find that the ratio of
the largest to the smallest parameters are 9.8, 8.0 and 220.1 for the
FSUGZ00, FSUGZ03 and FSUGZ06 parametrizations, respectively.  Though,
the FSUGZ03 parameter set is favoured from the ``naturalness" point
of view, it requires further improvement to obey the ``naturalness"
criteria in a more stringent manner. From the naturalness viewpoint,
FUGZ03 parametrization can be improved by considering  few additional
terms in the energy density functional (Eq.(\ref{eq:eden})).  Indeed,
it is clear from the findings of Ref.  \cite{Furnstahl97} that the next
higher order term containing gradients of the fields are necessary for the
overall improvements of the ``naturalness" behaviour of the parameters.

Before embarking on the discussion of our results, we would like to
focus on the strong correlations existing between the parameters
of the ERMF model which can be seen  from Tables \ref{tab:pfit} and
\ref{tab:dim_par}. It is evident from Eqs. (\ref{eq:cs2}) - (\ref{eq:v}) that
for given values of $\epsilon$, $K$, $M^*$ and $\rho_0$, the coupling
constants $g_\sigma$, $g_\omega$, $\overline\kappa$ and $\overline\lambda$ are
correlated with $\overline\alpha_1$, $\overline\alpha_1^\prime$ and $\zeta$. The
value of $\zeta$ also affects the symmetry energy coefficient and
its density dependence due to the mixed term containing $\omega$ and
$\rho$ meson fields. In other words, the coupling constants $g_\rho$ and
$\overline\alpha_3^\prime$ are also  correlated with $\zeta$. So,
it is very important to constrain the value of $\zeta$. This can
be achieved only if the behaviour of the EOS at higher densities
or the maximum mass of the neutron stars are known.  To constrain
the values of $\overline\alpha_1$ and $\overline\alpha_1^\prime$
one needs to know the density dependence of the effective meson
masses. The values of $\overline\alpha_2$, $\overline\alpha_2^\prime$ and
$\overline\alpha_3^\prime$ can be fixed by the neutron-skin thicknesses
of asymmetric nuclei.

\section{Nuclear and neutron matters}
\label{sec:snm_pnm}
 In this section we discuss our results for the symmetric nuclear matter
(SNM) and pure neutron matter (PNM). In Table
\ref{tab:nm_pro} we present our results for the various properties
associated with the SNM at the saturation density.
The quantity $L$ in Table \ref{tab:nm_pro} determines the linear density dependence of the
symmetry energy coefficient $J$ and is given by,
\begin{equation}
\label{eq:slp}
L = 3\rho \left . \frac{dJ}{d\rho}\right |_{\rho = \rho_0} 
\end{equation}
All the nuclear matter quantities given in the table for the FSUGZ and FSUGold parameters are
more or less close to each other. The values of nuclear matter
incompressibility coefficient, the symmetry energy coefficient and the
linear density dependence of symmetry energy  for the NL3 parameter
set are much higher than that for our new parameter sets. 

We use our newly generated parameter sets to study the effects of OMSC
 parameter  $\zeta$ on the  
EOS for the SNM and PNM.
 In
Fig. \ref{fig:del_ea} we plot the difference between the energy per
nucleon $\Delta\epsilon = \epsilon(\zeta=0)-\epsilon(\zeta)$ as a function
of $\zeta$ for SNM(dotted lines) and PNM(solid lines) at fixed densities
$\rho = 0.5$ fm$^{-3}$ and  0.8 fm$^{-3}$. We calculate 
$\epsilon (\zeta)$ at $\zeta = 0.0, 0.01,....,0.06$  using the
parameter sets FSUGZ00, FSUGZ01,...., FSUGZ06, respectively.  We see
 that $\Delta\epsilon$ increases with increase in $\zeta$. In other words,
the $\epsilon (\zeta)$ decreases with increase in $\zeta$ implying that EOS
becomes softer as $\zeta$ increases.  At higher densities the
values of $\epsilon$ appears to be very much sensitive to the value of
$\zeta$. For example, the value of $\Delta \epsilon$ at $\zeta =0.06$ is
less than 0.1MeV for $\rho = 0.2$ fm$^{-3}$ and it becomes greater than  125 MeV for $\rho
=0.8$ fm$^{-3}$. 
It must be pointed once again, 
 for different values of $\zeta$ the remaining parameters of the
ERMF models are so adjusted that they appropriately fit the properties of
the finite nuclei. 
 We find that FSUGZ03 parametrization  which corresponds to a moderate
value of $\zeta$
yields an EOS which is neither too stiff nor too soft. In Figs.
\ref{fig:snm_pnm}(a) and \ref{fig:snm_pnm}(b) we plot the EOS for SNM and PNM obtained using FSUGZ03
parametrization which
 agree  reasonably  well with the EOS
obtained within the DBHF framework \cite{Li92}.  The NL3 and FSUGold
parametrizations  yield the EOS which show larger deviations, in
particular, at higher densities  as compare
to the DBHF results. The NL3 parameters gives very stiff EOS while the
FSUGold parameters gives rise to very soft EOS.

\section{Finite nuclei}
\label{sec:fn}

The fit to the properties of the finite nuclei for the parameter sets
FSUGZ00 - FSUGZ06 are very much similar.
 For the
sake of clarity we present our results obtained for the finite nuclei
only for the parameter set FSUGZ03. In Fig.\ref{fig:be} we display
the values of the relative error in the total binding energies $
\delta B =(B^{exp}-B^{th})/B^{exp}$ in percent obtained for the FSUGZ03
parametrization. For the comparison, we also display the similar results
obtained for the NL3 and FSUGold parameter sets.  It is clear that the
binding energies obtained using FSUGZ03 parameter set agree better with the
experimental data than
those for the NL3 and FSUGold parameter sets.  The rms errors
 in the total binding energies are 1.6, 2.5 and 3.6 MeV calculated using
 FSUGZ03, NL3 and
FSUGold parameter sets, respectively. For other FSUGZ parameter sets the
rms errors in the
total binding energies lie in  between $1.5 - 1.8$ MeV.  In Fig. \ref{fig:rch} we display our results
for the relative error $\delta r_{ch}$  in percent for the charge rms
radii and compare them with the ones obtained for the NL3 and FSUGold
parametrizations. The values of charge rms radii calculated for various
parametrizations differ only marginally.
 The rms errors in the charge rms radii are 0.03,
0.02 and 0.03 fm for the FSUGZ03, NL3 and FSUGold parameter sets,
respectively.
Our results for the neutron-skin thickness $\Delta r = r_n - r_p$ are
plotted in Fig.  \ref{fig:nskin}. The  values of the rms radii $r_n$
and $r_p$  are calculated using the point density distributions for
neutrons and protons, respectively.  The values of $\Delta r$ shown by
crosses are the predictions based upon the 
 DBHF calculations \cite{Alonso03} and isospin diffusion data
\cite{Chen05}.  We see that the values of
$\Delta r$ obtained using FSUGZ03 and that for the FSUGold
parameter sets are closer to the predicted values. The NL3 parameters
significantly overestimates the values of $\Delta r$. We would like to
mention here that for the NL3 type of parametrization the values of the
symmetry energy coefficient $J$ and its linear density dependent $L$
are determined by a  single  parameter $g_\rho$ which determines the
coupling of $\rho$ mesons with the nucleons.  It may  be therefore  not
possible with NL3 type of  parametrization to fit simultaneously the values
of the binding energies and the neutron-skin thicknesses for asymmetric
nuclei. However, in the ERMF model, the mixed interactions of $\sigma$
and $\omega$ mesons with the $\rho$ mesons makes it possible to  give 
a wide spread in the values of neutron-skin thicknesses  
without affecting the other properties of finite nuclei
\cite{Furnstahl02,Sil05}. 

\section{Neutron stars}
\label{sec:ns}

In Sec. \ref{sec:snm_pnm} we have seen that the energy per
nucleon decreases as the OMSC parameter $\zeta$ increases (see
Fig.\ref{fig:del_ea}). This effect is more pronounced at higher densities
which implies that as $\zeta$ increases the EOS at higher densities
becomes softer.  Thus, it is natural to expect the differences  in the
properties of the neutron stars obtained for our new parameter sets with
different values of OMSC. In this section we present our results for the
properties of the neutron stars obtained for the seven different parameter
sets FSUGZ00 $-$ FSUGZ06.  The neutron star properties we have considered
are the limiting mass, central density and radius for the neutron star
with the ``canonical" mass of $1.4{\it M}_\odot$. These properties are
determined by integrating the  Tolman-Oppenheimer-Volkoff (TOV) equations
\cite{Weinberg72}. The TOV equations are solved using EOS for the matter
consisting of neutrons, protons, electrons and muons.  The  composition of
matter at any density is so determined that charge neutrality and $\beta
-$ equilibrium conditions are satisfied. For densities higher than $0.08$
fm$^{-3}$, the nuclear part of EOS is evaluated within the ERMF model
and that for electrons and muons the Fermi gas approximation is used.
At densities lower than $0.08$ fm$^{-3}$ down to $6.0\times 10^{-12}$
fm$^{-3}$ we use the EOS of Baym-Pethick-Sutherland  \cite{Baym71}.

In Fig. \ref{m-r} we plot mass-radius relationship for the neutron stars
calculated using FSUGZ00, FSUGZ03 and FSUGZ06 parametrizations.  
For the sake of comparison we also plot the similar results obtained
using FSUGold and NL3 parameter sets. The region bounded by $R\leqslant
3GM/c^2$ is excluded by the causality limit \cite{Lattimer90}. The line
labeled by $\Delta I/I = 0.014$ is radius limit estimated by Vela pulsar
glitches \cite{Lattimer01}.  The rotation constraint as indicated in Fig.
\ref{m-r} is obtained using \cite{Lattimer04},
\begin{equation}
\nu_k = 1045\left (\frac{\text M}{\text M_\odot}\right )^{1/2}\left ( \frac{10
\text km}{R}\right )^{3/2} \text Hz
\end{equation}
where, the frequency $\nu_k$ is taken to be 641 Hz  which is the highest
observed spin rate from the pulsar PSR B1937+21.
We see from Fig. \ref{m-r} that as the mass of the neutron stars
increases, the radius show stronger $\zeta$ dependence.  In particular,
radius for the $1.4{\it M}_\odot$ neutron stars decreases linearly
from 13.2 Km at $\zeta = 0.0$ to 12.4 Km at $\zeta = 0.06$.
We also calculate the radiation radius,
\begin{equation}
R_\infty=\frac{R}{\sqrt{1-\frac{2GM}{Rc^2}}}
\end{equation}
for the neutron with the canonical mass $1.4M_\odot$. For our parameter
sets, $R_\infty$ lies in the range of $15.19 - 15.93$ km.

In Fig.\ref{fig:zeta_mmax} we have plotted the variation of the
limiting mass ${\it M}_{max}$ for the neutron stars as a function of
OMSC parameter $\zeta$.   The $\zeta$ dependence of the various neutron
star properties considered here are obtained using the parameter sets
FSUGZ00 $-$ FSUGZ06 for which $\zeta$ takes the values $0.0 - 0.06$,
respectively. The remaining parameters for the FSUGZ sets are adjusted
to fit the properties of the finite nuclei. The symbols open circle and
square at $\zeta = 0.0$ and $0.06$ represent the values of the limiting
mass for the NL3 and FSUGold parametrizations, respectively.  The  value
of ${\it M}_{max}$ decreases with increase in $\zeta$, because, with the
increase in $\zeta$ the EOS becomes softer.  It is to be noted that the
NL3 and FSUGZ00 parametrizations have $\zeta = 0$. The only difference is
that the FSUGZ00 parameter set contains additional contributions from the
mixed interactions between $\sigma$, $\omega$ and $\rho$ mesons and values
of K, J and L  are much smaller.  The value of ${\it M}_{max}$ for FSUGold
and FSUGZ06 parameter sets are vary close, since, both these parameter
sets have $\zeta = 0.06$ and have similar nuclear matter properties.
For our case the ${\it M}_{max}$ varies between $2.3 {\it M}_\odot - 1.7
{\it M}_\odot$ for $0\leqslant \zeta \leqslant 0.06$. For the similar
range of $\zeta$, the calculations performed in Ref.\cite{Muller96}
yield the ${\it M}_{max}$ varying between $2.8 {\it M}_\odot - 1.8 {\it
M}_\odot$ which is significantly larger. The value of ${\it M}_{max}$
in Ref.\cite{Muller96} is obtained without including the contributions
from the mixed interactions. Unlike, in our case, the parameter sets for
different values of OMSC used in Ref.\cite{Muller96} are generated without
constraining the value of L as given by Eq.(\ref{eq:slp}).  We must also
point out that for moderate values of OMSC ($\zeta\sim 0.03$) we get
${\it M}_{max}$ $\sim$ 1.9 ${\it M}_\odot$.  In the present work we do not
consider the influence of the self-coupling of the $\rho$ mesons on the
maximum mass of the neutron stars. It is shown in Ref. \cite{Muller96}
that when the $\rho$ meson self-coupling is varied within the bounds of
the ``naturalness", the maximum mass of the neutron star  composed of
the $\beta$-equilibrated matter can increase at most by 0.1$M_\odot$.

In Fig. \ref{yp_rho} we plot our results for the proton fractions as a
function of baryon density. The solid circles represent the threshold
density at which the condition \cite{Lattimer91},
\begin{equation}
Y_n^{1/3} \leqslant Y_p^{1/3} + Y_e^{1/3}
\label{eq:urca}
\end{equation}
for the direct Urca (DU)  process to occur is satisfied, where, $Y_i =
\rho_i/\rho$ is the fraction for the $i-$th  species. 
For the neutron star composed of neutrons, protons, electrons and
muons, the critical value of the proton fraction $Y_{DU}$ at which DU
process sets in can be obtained from Eq. (\ref{eq:urca}) together with the charge neutrality condition as,
\begin{equation}
Y_{DU} = \frac{1}
{1+\left [ 1+\left (\frac{Y_e}{Y_e+Y_\mu}\right)^{1/3}\right]^3} .
\label{eq:ydu}
\end{equation}
It is clear from the Eq. (\ref{eq:ydu}) that at very low densities when
$Y_\mu = 0$, $Y_{DU}$ becomes 0.11.  At extremely high densities when
$Y_e\approx Y_\mu = \frac{1}{2}$, $Y_{DU}$ becomes  0.15.  From Fig.
\ref{yp_rho}
we see that the critical proton fraction is $\sim 0.13$.
In Fig.\ref{fig:zeta_rhouc} we  plot the  variations of the densities
$\rho_{\scriptstyle{U}}$ and $\rho_c$ as a function of $\zeta$, where, the
$\rho_{\scriptstyle{U}}$ is the density at which the direct Urca process
sets in and   $\rho_c$   denotes the central density for the neutron
star with the ``canonical" mass of $1.4{\it M}_\odot$. As expected,
$\rho_{\scriptstyle{U}}$ and $\rho_c$ increase with $\zeta$. However, it
is interesting to note that  $\rho_{\scriptstyle {U}}$ $>$ $\rho_c$ for
smaller $\zeta$ and $\rho_{\scriptstyle{U}} < \rho_c$ for $\zeta \gtrsim
0.02$. This means that the direct Urca process can occur in the neutron
stars for the ``canonical" mass of $1.4 {\it M}_\odot$ only for $\zeta
\gtrsim 0.02$.  Also, one can infer from the Fig.\ref{fig:zeta_rhouc} that
the minimum  mass ${\it M}_U$ for the neutron star in which  direct Urca
process can occur will decrease with $\zeta$. In Fig. \ref{fig:zeta_mu}
we plot ${\it M}_U$ versus $\zeta$.

We have not considered the hyperonic degrees of freedom in our
calculations. The EOS with nucleons, hyperons and leptons are softer
than the one without hyperons \cite{Glendenning00}.  The stiffer is
the EOS without hyperons, more is the softening effect when hyperons
are included \cite{Balberg99,Schulze06}. Thus, the results obtained with
smaller values of $\zeta$ will be affected more than the ones obtained
for larger values of $\zeta$. As a result, we expect that the variations
in the maximum mass with $\zeta$ would certainly decrease.  Further,
in the presence of hyperons the threshold density  for the direct Urca
process will be pushed up.

\newpage
\section{Conclusions}
\label{sec:conc}

The effects of OMSC on the properties of finite nuclei and neutron stars
are investigated using a ERMF model which includes  the contributions
from all possible  mixed interactions between the $\sigma$, $\omega$ and
$\rho$ mesons  upto the quartic order.  The simulated annealing method is
implemented for minimizing the $\chi ^2$ function required to determine
the best fit parameters. For a realistic  investigation, we generate
seven different parameter sets named FSUGZ00, FSUGZ01,...., FSUGZ06
for  different values of OMSC by adjusting  the remaining parameters of
the model to fit the properties of the finite nuclei. The properties
of finite nuclei used in the fits are  the total binding energies and
charge rms radii for several closed shell nuclei ranging from normal
to exotic ones. In addition, we also include in our fit  neutron-skin
thickness for $^{208}$Pb nucleus to constrain the density dependence of
the symmetry energy coefficient. The value of neutron-skin thickness is
so chosen that it agrees reasonably well with the recent predictions
\cite{Piekarewicz04,Alonso03,Chen05}.  All these parameter sets  fit
equally well the finite nuclear properties. But, only moderate values
of OMSC are favored from the ``naturalness" point of view.
For our parameter sets  the rms errors in the total binding
energies calculated for the nuclei used in the fits are $1.5 - 1.8$
MeV which are significantly smaller than $3.6$ MeV obtained for recently
proposed FSUGold parametrization.

The behaviour of  EOS at higher densities is predominantly determined
by the OMSC which is different for our different parameter sets.
The parameter sets with  moderate values of the OMSC not  only exhibit
the ``naturalness", but, also yield the EOS for the symmetric nuclear
and pure neutron matters which closely resemble the ones calculated
within the framework of the Dirac-Brueckner- Hartree-Fock.  For such
parameter sets the values of the limiting mass for the neutron stars
is $\sim 1.9{\it M}_\odot$.  For our various parametrizations the
value of the limiting mass of the neutrons stars lie in the range of
$1.7{\it M}_\odot - 2.3{\it M}_\odot$. We also find that the direct Urca
process can occur in the neutron stars with ``canonical" mass of $1.4
{\it M}_\odot$ only for the moderate and higher values of OMSC. For our
various parametrizations, the radius of $1.4{\it M}_\odot$ neutron stars
lie in the range of 12.4 - 13.2 Km.

\begin{acknowledgments}
This work was supported in part by the University Grant Commission under grant
\# F.17-40/98 (SA-I).
\end{acknowledgments}
\newpage
% \bibliography{1review}

\newpage
\begin{thebibliography}{38}
\expandafter\ifx\csname natexlab\endcsname\relax\def\natexlab#1{#1}\fi
\expandafter\ifx\csname bibnamefont\endcsname\relax
  \def\bibnamefont#1{#1}\fi
\expandafter\ifx\csname bibfnamefont\endcsname\relax
  \def\bibfnamefont#1{#1}\fi
\expandafter\ifx\csname citenamefont\endcsname\relax
  \def\citenamefont#1{#1}\fi
\expandafter\ifx\csname url\endcsname\relax
  \def\url#1{\texttt{#1}}\fi
\expandafter\ifx\csname urlprefix\endcsname\relax\def\urlprefix{URL }\fi
\providecommand{\bibinfo}[2]{#2}
\providecommand{\eprint}[2][]{\url{#2}}

\bibitem[{\citenamefont{Furnstahl et~al.}(1996)\citenamefont{Furnstahl, Serot,
  and Tang}}]{Furnstahl96}
\bibinfo{author}{\bibfnamefont{R.}~\bibnamefont{Furnstahl}},
  \bibinfo{author}{\bibfnamefont{B.~D.} \bibnamefont{Serot}}, \bibnamefont{and}
  \bibinfo{author}{\bibfnamefont{H.-B.} \bibnamefont{Tang}},
  \bibinfo{journal}{Nucl. Phys.} \textbf{\bibinfo{volume}{A598}},
  \bibinfo{pages}{539} (\bibinfo{year}{1996}).

\bibitem[{\citenamefont{Serot and Walecka}(1997)}]{Serot97}
\bibinfo{author}{\bibfnamefont{B.~D.} \bibnamefont{Serot}} \bibnamefont{and}
  \bibinfo{author}{\bibfnamefont{J.~D.} \bibnamefont{Walecka}},
  \bibinfo{journal}{Int. J. Mod. Phys. E} \textbf{\bibinfo{volume}{6}},
  \bibinfo{pages}{515} (\bibinfo{year}{1997}).

\bibitem[{\citenamefont{Furnstahl et~al.}(1997)\citenamefont{Furnstahl, Serot,
  and Tang}}]{Furnstahl97}
\bibinfo{author}{\bibfnamefont{R.}~\bibnamefont{Furnstahl}},
  \bibinfo{author}{\bibfnamefont{B.~D.} \bibnamefont{Serot}}, \bibnamefont{and}
  \bibinfo{author}{\bibfnamefont{H.-B.} \bibnamefont{Tang}},
  \bibinfo{journal}{Nucl. Phys.} \textbf{\bibinfo{volume}{A615}},
  \bibinfo{pages}{441} (\bibinfo{year}{1997}).

\bibitem[{\citenamefont{Walecka}(1974)}]{Walecka74}
\bibinfo{author}{\bibfnamefont{J.~D.} \bibnamefont{Walecka}},
  \bibinfo{journal}{Ann. Phys. (N.Y.)} \textbf{\bibinfo{volume}{83}},
  \bibinfo{pages}{491} (\bibinfo{year}{1974}).

\bibitem[{\citenamefont{Boguta and Bodmer}(1977)}]{Boguta77}
\bibinfo{author}{\bibfnamefont{J.}~\bibnamefont{Boguta}} \bibnamefont{and}
  \bibinfo{author}{\bibfnamefont{A.~R.} \bibnamefont{Bodmer}},
  \bibinfo{journal}{Nucl. Phys.} \textbf{\bibinfo{volume}{A292}},
  \bibinfo{pages}{413} (\bibinfo{year}{1977}).

\bibitem[{\citenamefont{M{\"u}ller and Serot}(1996)}]{Muller96}
\bibinfo{author}{\bibfnamefont{H.}~\bibnamefont{M{\"u}ller}} \bibnamefont{and}
  \bibinfo{author}{\bibfnamefont{B.~D.} \bibnamefont{Serot}},
  \bibinfo{journal}{Nucl. Phys.} \textbf{\bibinfo{volume}{A606}},
  \bibinfo{pages}{508} (\bibinfo{year}{1996}).

\bibitem[{\citenamefont{Estal et~al.}(2001)\citenamefont{Estal, Centelles,
  Viñas, and Patra}}]{Estal01}
\bibinfo{author}{\bibfnamefont{M.~D.} \bibnamefont{Estal}},
  \bibinfo{author}{\bibfnamefont{M.}~\bibnamefont{Centelles}},
  \bibinfo{author}{\bibfnamefont{X.}~\bibnamefont{Viñas}}, \bibnamefont{and}
  \bibinfo{author}{\bibfnamefont{S.~K.} \bibnamefont{Patra}},
  \bibinfo{journal}{Phys. Rev. C} \textbf{\bibinfo{volume}{63}},
  \bibinfo{pages}{024314} (\bibinfo{year}{2001}).

\bibitem[{\citenamefont{Todd-Rutel and Piekarewicz}(2005)}]{Todd-Rutel05}
\bibinfo{author}{\bibfnamefont{B.~G.} \bibnamefont{Todd-Rutel}}
  \bibnamefont{and}
  \bibinfo{author}{\bibfnamefont{J.}~\bibnamefont{Piekarewicz}},
  \bibinfo{journal}{Phys. Rev. Lett} \textbf{\bibinfo{volume}{95}},
  \bibinfo{pages}{122501} (\bibinfo{year}{2005}).

\bibitem[{\citenamefont{Piekarewicz and Weppner}(2005)}]{Piekarewicz05}
\bibinfo{author}{\bibfnamefont{J.}~\bibnamefont{Piekarewicz}} \bibnamefont{and}
  \bibinfo{author}{\bibfnamefont{S.~P.} \bibnamefont{Weppner}},
  \bibinfo{journal}{nucl-th/0509019}  (\bibinfo{year}{2005}).

\bibitem[{\citenamefont{Glendenning and Moszkowski}(1991)}]{Glendenning91}
\bibinfo{author}{\bibfnamefont{N.~K.} \bibnamefont{Glendenning}}
  \bibnamefont{and} \bibinfo{author}{\bibfnamefont{S.~A.}
  \bibnamefont{Moszkowski}}, \bibinfo{journal}{Phys. Rev. Lett.}
  \textbf{\bibinfo{volume}{67}}, \bibinfo{pages}{2414} (\bibinfo{year}{1991}).

\bibitem[{\citenamefont{Ban et~al.}(2004)\citenamefont{Ban, Li, Zhang, Jia,
  Sang, and Meng}}]{Ban04}
\bibinfo{author}{\bibfnamefont{S.~F.} \bibnamefont{Ban}},
  \bibinfo{author}{\bibfnamefont{J.}~\bibnamefont{Li}},
  \bibinfo{author}{\bibfnamefont{S.~Q.} \bibnamefont{Zhang}},
  \bibinfo{author}{\bibfnamefont{H.~Y.} \bibnamefont{Jia}},
  \bibinfo{author}{\bibfnamefont{J.~P.} \bibnamefont{Sang}}, \bibnamefont{and}
  \bibinfo{author}{\bibfnamefont{J.}~\bibnamefont{Meng}},
  \bibinfo{journal}{Phys. Rev. C} \textbf{\bibinfo{volume}{69}},
  \bibinfo{pages}{045805} (\bibinfo{year}{2004}).

\bibitem[{\citenamefont{Agrawal et~al.}(2005)\citenamefont{Agrawal, Shlomo, and
  Au}}]{Agrawal05}
\bibinfo{author}{\bibfnamefont{B.~K.} \bibnamefont{Agrawal}},
  \bibinfo{author}{\bibfnamefont{S.}~\bibnamefont{Shlomo}}, \bibnamefont{and}
  \bibinfo{author}{\bibfnamefont{V.~K.} \bibnamefont{Au}},
  \bibinfo{journal}{Phys. Rev. C} \textbf{\bibinfo{volume}{72}},
  \bibinfo{pages}{014310} (\bibinfo{year}{2005}).

\bibitem[{\citenamefont{Agrawal et~al.}(2006)\citenamefont{Agrawal, Dhiman, and
  Kumar}}]{Agrawal06}
\bibinfo{author}{\bibfnamefont{B.~K.} \bibnamefont{Agrawal}},
  \bibinfo{author}{\bibfnamefont{S.~K.} \bibnamefont{Dhiman}},
  \bibnamefont{and} \bibinfo{author}{\bibfnamefont{R.}~\bibnamefont{Kumar}},
  \bibinfo{journal}{Phys. Rev. C} \textbf{\bibinfo{volume}{73}},
  \bibinfo{pages}{034319} (\bibinfo{year}{2006}).

\bibitem[{\citenamefont{Lalazissis et~al.}(1997)\citenamefont{Lalazissis,
  Konig, and Ring}}]{Lalazissis97}
\bibinfo{author}{\bibfnamefont{G.~A.} \bibnamefont{Lalazissis}},
  \bibinfo{author}{\bibfnamefont{J.}~\bibnamefont{Konig}}, \bibnamefont{and}
  \bibinfo{author}{\bibfnamefont{P.}~\bibnamefont{Ring}},
  \bibinfo{journal}{Phys. Rev. C} \textbf{\bibinfo{volume}{55}},
  \bibinfo{pages}{540} (\bibinfo{year}{1997}).

\bibitem[{\citenamefont{Kirkpatrik}(1984)}]{Patrik84}
\bibinfo{author}{\bibfnamefont{S.}~\bibnamefont{Kirkpatrik}},
  \bibinfo{journal}{J. Stat. Phys.} \textbf{\bibinfo{volume}{34}},
  \bibinfo{pages}{975} (\bibinfo{year}{1984}).

\bibitem[{\citenamefont{Ingber}(1989)}]{Ingber89}
\bibinfo{author}{\bibfnamefont{L.}~\bibnamefont{Ingber}},
  \bibinfo{journal}{Mathl. Comput. Modelling} \textbf{\bibinfo{volume}{12}},
  \bibinfo{pages}{967} (\bibinfo{year}{1989}).

\bibitem[{\citenamefont{Cohen}(1994)}]{Cohen94}
\bibinfo{author}{\bibfnamefont{B.}~\bibnamefont{Cohen}}, Master's thesis,
  \bibinfo{school}{Tel-Aviv University} (\bibinfo{year}{1994}).

\bibitem[{\citenamefont{B{\"u}rvenich et~al.}(2002)\citenamefont{B{\"u}rvenich,
  Madland, Maruhn, and Reinhard}}]{Burvenich02}
\bibinfo{author}{\bibfnamefont{T.}~\bibnamefont{B{\"u}rvenich}},
  \bibinfo{author}{\bibfnamefont{D.~G.} \bibnamefont{Madland}},
  \bibinfo{author}{\bibfnamefont{J.~A.} \bibnamefont{Maruhn}},
  \bibnamefont{and} \bibinfo{author}{\bibfnamefont{P.-G.}
  \bibnamefont{Reinhard}}, \bibinfo{journal}{Phys. Rev. C}
  \textbf{\bibinfo{volume}{65}}, \bibinfo{pages}{044308}
  (\bibinfo{year}{2002}).

\bibitem[{\citenamefont{B{\"u}rvenich et~al.}(2004)\citenamefont{B{\"u}rvenich,
  Madland, and Reinhard}}]{Burvenich04}
\bibinfo{author}{\bibfnamefont{T.}~\bibnamefont{B{\"u}rvenich}},
  \bibinfo{author}{\bibfnamefont{D.~G.} \bibnamefont{Madland}},
  \bibnamefont{and} \bibinfo{author}{\bibfnamefont{P.-G.}
  \bibnamefont{Reinhard}}, \bibinfo{journal}{Nucl. Phys.}
  \textbf{\bibinfo{volume}{A744}}, \bibinfo{pages}{92} (\bibinfo{year}{2004}).

\bibitem[{\citenamefont{Bodmer}(1991)}]{Bodmer91}
\bibinfo{author}{\bibfnamefont{A.~R.} \bibnamefont{Bodmer}},
  \bibinfo{journal}{Nucl. Phys.} \textbf{\bibinfo{volume}{A526}},
  \bibinfo{pages}{703} (\bibinfo{year}{1991}).

\bibitem[{\citenamefont{Audi et~al.}(2003)\citenamefont{Audi, Wapstra, and
  Thibault}}]{Audi03}
\bibinfo{author}{\bibfnamefont{G.}~\bibnamefont{Audi}},
  \bibinfo{author}{\bibfnamefont{A.~H.} \bibnamefont{Wapstra}},
  \bibnamefont{and} \bibinfo{author}{\bibfnamefont{C.}~\bibnamefont{Thibault}},
  \bibinfo{journal}{Nucl. Phys.} \textbf{\bibinfo{volume}{A729}},
  \bibinfo{pages}{337} (\bibinfo{year}{2003}).

\bibitem[{\citenamefont{Otten}(1989)}]{Otten89}
\bibinfo{author}{\bibfnamefont{E.~W.} \bibnamefont{Otten}},
  \emph{\bibinfo{title}{in Treatise on Heavy-Ion Science}},
  vol.~\bibinfo{volume}{8} (\bibinfo{publisher}{ed. D. A. Bromley, Plenum, New
  York}, \bibinfo{year}{1989}).

\bibitem[{\citenamefont{Vries et~al.}(1987)\citenamefont{Vries, Jager, and
  Vries}}]{Vries87}
\bibinfo{author}{\bibfnamefont{H.~D.} \bibnamefont{Vries}},
  \bibinfo{author}{\bibfnamefont{C.~W.~D.} \bibnamefont{Jager}},
  \bibnamefont{and} \bibinfo{author}{\bibfnamefont{C.~D.} \bibnamefont{Vries}},
  \bibinfo{journal}{At. Data Nucl. Data Tables} \textbf{\bibinfo{volume}{36}},
  \bibinfo{pages}{495} (\bibinfo{year}{1987}).

\bibitem[{\citenamefont{Piekarewicz}(2004)}]{Piekarewicz04}
\bibinfo{author}{\bibfnamefont{J.}~\bibnamefont{Piekarewicz}},
  \bibinfo{journal}{Phys. Rev. C} \textbf{\bibinfo{volume}{69}},
  \bibinfo{pages}{041301} (\bibinfo{year}{2004}).

\bibitem[{\citenamefont{Alonso and Sammarruca}(2003)}]{Alonso03}
\bibinfo{author}{\bibfnamefont{D.}~\bibnamefont{Alonso}} \bibnamefont{and}
  \bibinfo{author}{\bibfnamefont{F.}~\bibnamefont{Sammarruca}},
  \bibinfo{journal}{Phys. Rev. C} \textbf{\bibinfo{volume}{68}},
  \bibinfo{pages}{054305} (\bibinfo{year}{2003}).

\bibitem[{\citenamefont{Chen et~al.}(2005)\citenamefont{Chen, Ko, and
  Li}}]{Chen05}
\bibinfo{author}{\bibfnamefont{L.-W.} \bibnamefont{Chen}},
  \bibinfo{author}{\bibfnamefont{C.~M.} \bibnamefont{Ko}}, \bibnamefont{and}
  \bibinfo{author}{\bibfnamefont{B.-A.} \bibnamefont{Li}},
  \bibinfo{journal}{Phys. Rev. C} \textbf{\bibinfo{volume}{72}},
  \bibinfo{pages}{064309} (\bibinfo{year}{2005}).

\bibitem[{\citenamefont{Li et~al.}(1992)\citenamefont{Li, Machleidt, and
  Brockmann}}]{Li92}
\bibinfo{author}{\bibfnamefont{G.~Q.} \bibnamefont{Li}},
  \bibinfo{author}{\bibfnamefont{R.}~\bibnamefont{Machleidt}},
  \bibnamefont{and}
  \bibinfo{author}{\bibfnamefont{R.}~\bibnamefont{Brockmann}},
  \bibinfo{journal}{Phys. Rev. C} \textbf{\bibinfo{volume}{45}},
  \bibinfo{pages}{2782} (\bibinfo{year}{1992}).

\bibitem[{\citenamefont{Furnstahl}(2002)}]{Furnstahl02}
\bibinfo{author}{\bibfnamefont{R.}~\bibnamefont{Furnstahl}},
  \bibinfo{journal}{Nucl. Phys.} \textbf{\bibinfo{volume}{A706}},
  \bibinfo{pages}{85} (\bibinfo{year}{2002}).

\bibitem[{\citenamefont{Sil et~al.}(2005)\citenamefont{Sil, Centelles, Vinas,
  and Piekarewicz}}]{Sil05}
\bibinfo{author}{\bibfnamefont{T.}~\bibnamefont{Sil}},
  \bibinfo{author}{\bibfnamefont{M.}~\bibnamefont{Centelles}},
  \bibinfo{author}{\bibfnamefont{X.}~\bibnamefont{Vinas}}, \bibnamefont{and}
  \bibinfo{author}{\bibfnamefont{J.}~\bibnamefont{Piekarewicz}},
  \bibinfo{journal}{Phys. Rev.} \textbf{\bibinfo{volume}{C71}},
  \bibinfo{pages}{045502} (\bibinfo{year}{2005}).

\bibitem[{\citenamefont{Weinberg}(1972)}]{Weinberg72}
\bibinfo{author}{\bibfnamefont{S.}~\bibnamefont{Weinberg}},
  \emph{\bibinfo{title}{Gravitation and cosmology}} (\bibinfo{publisher}{Wiley,
  New York}, \bibinfo{year}{1972}).

\bibitem[{\citenamefont{Baym et~al.}(1971)\citenamefont{Baym, Pethick, and
  Sutherland}}]{Baym71}
\bibinfo{author}{\bibfnamefont{G.}~\bibnamefont{Baym}},
  \bibinfo{author}{\bibfnamefont{C.}~\bibnamefont{Pethick}}, \bibnamefont{and}
  \bibinfo{author}{\bibfnamefont{P.}~\bibnamefont{Sutherland}},
  \bibinfo{journal}{Astrophys. J.} \textbf{\bibinfo{volume}{170}},
  \bibinfo{pages}{299} (\bibinfo{year}{1971}).

\bibitem[{\citenamefont{Lattimer et~al.}(1990)\citenamefont{Lattimer, Prakash,
  Masak, and Yahil}}]{Lattimer90}
\bibinfo{author}{\bibfnamefont{J.~M.} \bibnamefont{Lattimer}},
  \bibinfo{author}{\bibfnamefont{M.}~\bibnamefont{Prakash}},
  \bibinfo{author}{\bibfnamefont{D.}~\bibnamefont{Masak}}, \bibnamefont{and}
  \bibinfo{author}{\bibfnamefont{A.}~\bibnamefont{Yahil}},
  \bibinfo{journal}{Astrophys. J.} \textbf{\bibinfo{volume}{355}},
  \bibinfo{pages}{241} (\bibinfo{year}{1990}).

\bibitem[{\citenamefont{Lattimer and Prakash}(2001)}]{Lattimer01}
\bibinfo{author}{\bibfnamefont{J.~M.} \bibnamefont{Lattimer}} \bibnamefont{and}
  \bibinfo{author}{\bibfnamefont{M.}~\bibnamefont{Prakash}},
  \bibinfo{journal}{Astrophys. J.} \textbf{\bibinfo{volume}{550}},
  \bibinfo{pages}{426} (\bibinfo{year}{2001}).

\bibitem[{\citenamefont{Lattimer and Prakash}(2004)}]{Lattimer04}
\bibinfo{author}{\bibfnamefont{J.~M.} \bibnamefont{Lattimer}} \bibnamefont{and}
  \bibinfo{author}{\bibfnamefont{M.}~\bibnamefont{Prakash}},
  \bibinfo{journal}{Science} \textbf{\bibinfo{volume}{304}},
  \bibinfo{pages}{536} (\bibinfo{year}{2004}).

\bibitem[{\citenamefont{Lattimer et~al.}(1991)\citenamefont{Lattimer, Pethick,
  M.Prakash, and Haensel}}]{Lattimer91}
\bibinfo{author}{\bibfnamefont{J.~M.} \bibnamefont{Lattimer}},
  \bibinfo{author}{\bibfnamefont{C.~J.} \bibnamefont{Pethick}},
  \bibinfo{author}{\bibnamefont{M.Prakash}}, \bibnamefont{and}
  \bibinfo{author}{\bibfnamefont{P.}~\bibnamefont{Haensel}},
  \bibinfo{journal}{Phys. Rev. Lett.} \textbf{\bibinfo{volume}{66}},
  \bibinfo{pages}{2701} (\bibinfo{year}{1991}).

\bibitem[{\citenamefont{Glendenning}(2000)}]{Glendenning00}
\bibinfo{author}{\bibfnamefont{N.}~\bibnamefont{Glendenning}},
  \emph{\bibinfo{title}{Compect Stars: Nuclear Physics, Particle Physics, and
  General Relativity}} (\bibinfo{publisher}{Springer, New York},
  \bibinfo{year}{2000}).

\bibitem[{\citenamefont{Balberg et~al.}(1999)\citenamefont{Balberg,
  Lichtenstadt, and B.Cook}}]{Balberg99}
\bibinfo{author}{\bibfnamefont{S.}~\bibnamefont{Balberg}},
  \bibinfo{author}{\bibfnamefont{I.}~\bibnamefont{Lichtenstadt}},
  \bibnamefont{and} \bibinfo{author}{\bibfnamefont{G.}~\bibnamefont{B.Cook}},
  \bibinfo{journal}{Astrophys. J.} \textbf{\bibinfo{volume}{S121}},
  \bibinfo{pages}{515} (\bibinfo{year}{1999}).

\bibitem[{\citenamefont{Schulze et~al.}(2006)\citenamefont{Schulze, Polls,
  Ramos, and Vidana}}]{Schulze06}
\bibinfo{author}{\bibfnamefont{H.~J.} \bibnamefont{Schulze}},
  \bibinfo{author}{\bibfnamefont{A.}~\bibnamefont{Polls}},
  \bibinfo{author}{\bibfnamefont{A.}~\bibnamefont{Ramos}}, \bibnamefont{and}
  \bibinfo{author}{\bibfnamefont{I.}~\bibnamefont{Vidana}},
  \bibinfo{journal}{Phys. Rev. C} \textbf{\bibinfo{volume}{73}},
  \bibinfo{pages}{058801} (\bibinfo{year}{2006}).

\end{thebibliography}

\newpage
 \begin{table}[p]
 \caption{\label{tab:prange} The lower (${\bf
v_0}$) and upper (${\bf v_1}$)  limits, maximum displacement (${\bf d}$)
and initial values (${\bf v}_{in}$)  for the ERMF model parameters used
for minimizing the $\chi^2$ value within the SAM. The nucleon, omega and
rho masses are kept fixed at
 $M= 939$ MeV, $m_\omega = 782.5$ MeV and $m_\rho = 770$ MeV,
respectively.}

\begin{ruledtabular}
\begin{tabular}{|cdddd|}
\multicolumn{1}{|c}{}&
\multicolumn{1}{c}{${\bf v}_0$}&
\multicolumn{1}{c}{${\bf v}_1$}&
\multicolumn{1}{c}{${\bf d}$}&
\multicolumn{1}{c|}{${\bf v}_{in}$}\\
\hline
$\epsilon$(MeV)  & -16.50 &  -15.50 &  0.1& -16.00\\
$K$(MeV)&      220.0&  240.0 & 2.0&230.0\\
$M^*/M$  &0.60&  0.70 & 0.01&0.65\\
$\rho_0$(fm$^{-3}$) &0.145  & 0.165&  0.002&0.155\\
$g_\rho$ & 10.0  & 15.0 & 0.5&12.50\\
$\overline{\alpha}_1$(fm$^{-1}$)  & 0.0&  0.0095 & 0.0010&0.005\\
$\overline{\alpha}_1^\prime$ &0.0  &0.0040 & 0.0004&0.002\\
$\overline{\alpha}_2$(fm$^{-1}$)  &0.0  &0.0380 & 0.0038&0.019\\
$\overline{\alpha}_2^\prime$  &0.0  &0.0160 & 0.0016&0.008\\
$\overline{\alpha}_3^\prime$  & 0.0  &0.0160 & 0.0016& 0.008\\
$m_\sigma$(MeV) & 490.0&510.0&2.0 &500.0\\

\end{tabular}
\end{ruledtabular}
\end{table}

\begin{table}[p]
\caption{\label{tab:pfit} Various newly generated  parameter sets FSUGZ00,
FSUGZ03 and FSUGZ06 for the ERMF models. The
parameters $\overline{\kappa}$, $\overline{\alpha}_1$, $\overline{\alpha}_2$
are given in units of fm$^{-1}$.
The nucleon mass $M$ and meson masses  $m_\sigma$,  $m_\omega$ and  $m_\rho$ 
 are given in units of MeV. The parameters for the NL3 and FSUGold sets are
taken from Refs. \cite{Lalazissis97,Todd-Rutel05}}

\begin{ruledtabular}
\begin{tabular}{|cddddd|}
\multicolumn{1}{|c}{${\bf Parameters}$}&
\multicolumn{1}{c}{${\bf NL3}$}&
\multicolumn{1}{c}{${\bf FSUGold}$}&
\multicolumn{1}{c}{${\bf FSUGZ00}$}&
\multicolumn{1}{c}{${\bf FSUGZ03}$}&
\multicolumn{1}{c|}{${\bf FSUGZ06}$}\\
\hline
$g_\sigma$  & 10.21743 & 10.59217 &10.65616 &10.76145&11.02412\\
$g_\omega$ &12.86764  &  14.30207 &13.95799 &14.11104&14.66595\\
$g_\rho$ &8.94880  & 11.76733& 14.32687 &14.67414&14.52185\\
$\overline{\kappa}$  &0.019573&  0.007194&0.030215 &0.015606 &0.006949\\
$\overline{\lambda}$ & -0.015914  & 0.023762&-0.004544&0.009753 &0.024487\\
$\zeta$ & -  & 0.060000 &0.0&0.030000&0.060000\\
$\overline{\alpha}_1$ & -&  - &0.003867 &0.001031&0.000045\\
$\overline{\alpha}_1^\prime$ &- &-&0.000779&0.000507 &0.000053\\
$\overline{\alpha}_2$  &-  &- & 0.029179&0.030682&0.025836\\
$\overline{\alpha}_2^\prime$  &-  &-&0.013501&0.011625 & 0.015688\\
$\overline{\alpha}_3^\prime$  & -  &0.600000 &0.014759&0.013598&0.015849\\
$M$ & 939&939&939 &939&939\\
$m_\sigma$ & 508.194&491.500 &495.763&500.511&501.370 \\
$m_\omega$ & 782.501&782.500&782.500&782.500&782.500 \\
$m_\rho$ & 763.0&763.0&770.0 &770.0&770.0\\
\end{tabular}
\end{ruledtabular}
\end{table}

\begin{table}[p]
\caption{\label{tab:dim_par} The values of parameters expressed  as
dimensionless ratios using Eq.(\ref{eq:scl_par}). 
 All the values 
have been multiplied by $10^3$ }
\begin{ruledtabular}
\begin{tabular}{|cddddd|}
\multicolumn{1}{|c}{${}$}&
\multicolumn{1}{c}{${\bf NL3}$}&
\multicolumn{1}{c}{${\bf FSUGold}$}&
\multicolumn{1}{c}{${\bf FSUGZ00}$}&
\multicolumn{1}{c}{${\bf FSUGZ03}$}&
\multicolumn{1}{c|}{${\bf FSUGZ06}$}\\
\hline
$\frac{1}{2c_\sigma^2 M^2}$ &1.403  & 1.221 & 1.227&1.227&1.173\\
$\frac{1}{2c_\omega^2 M^2}$ &2.097  & 1.698 &1.782&1.744&1.614\\
$\frac{1}{8c_\rho^2 M^2}$  &1.030  & 0.596  & 0.409&0.390&0.399\\
$\frac{\overline{\kappa}}{6 M}$  & 0.686 & 0.252 &1.0508&0.547& 0.243\\
$\frac{\overline{\lambda}}{24}$  & -0.663  &0.990 &-0.189&0.406& 1.020\\
$\frac{\zeta}{24}$   & -  & 2.500  & -&1.250&2.500\\
$\frac{\overline{\alpha}_1}{M}$  & - & -&0.813&0.217 & 0.009\\
$\frac{\overline{\alpha}_1^\prime}{2}$  &- &-&0.389 & 0.253 &0.026\\
$\frac{\overline{\alpha}_2}{4 M}$ & - & - &1.533&1.612&1.357\\
$\frac{\overline{\alpha}_2^\prime}{8}$  & - &-&1.688& 1.453 & 1.961\\
$\frac{\overline{\alpha}_3^\prime}{8}$ & - &7.500&1.844&1.699 &1.981\\
\end{tabular}
\end{ruledtabular}
\end{table}

\begin{table}[p]
\caption{\label{tab:nm_pro} The Nuclear matter properties 
 at the saturation density  for newly generated parameter sets FSUGZ00, FSUGZ03 and FSUGZ06 are compared with
the corresponding ones obtained using NL3 and FSUGold parameter sets.
The quantities given below are: $\epsilon$ the binding energy per nucleon,
$K$ the nuclear matter incompressibility coefficient,
$J$ the symmetry energy, $L=3\rho\frac{dJ}{d\rho}$
related to the slope of the symmetry energy, $M^*/M$ is the ratio of the
 effective  nucleon  mass to the  nucleon mass and  $\rho_0$
the saturation density.}  

\begin{ruledtabular}
\begin{tabular}{|cccccc|}
\multicolumn{1}{|c}{}&
\multicolumn{1}{c}{${\bf NL3}$}&
\multicolumn{1}{c}{${\bf FSUGold}$}&
\multicolumn{1}{c}{${\bf FSUGZ00}$}&
\multicolumn{1}{c}{${\bf FSUGZ03}$}&
\multicolumn{1}{c|}{${\bf FSUGZ06}$}\\
\hline
$\epsilon$(MeV)  & -16.25 & -16.29&-16.03&-16.07& -16.05\\ 
$K$(MeV)&      271.5&  230.0 & 240.0&232.5&224.9\\
$J$ (MeV) & 37.40 & 32.59 & 31.43&31.55&31.17\\
$L$ (MeV) & 118.56 & 60.56 & 62.19&64.01&62.43\\
$M^*/M$  &0.595&  0.609 & 0.605&0.603&0.607\\
$\rho_0$(fm$^{-3}$) &0.148  & 0.148&0.149&0.147& 0.146\\
\end{tabular}
\end{ruledtabular}
\end{table}

\begin{figure}

\caption{\label{fig:del_ea} The difference between energy per nucleon
$\Delta \epsilon=\epsilon(\zeta=0)-\epsilon(\zeta)$ are plotted as a
function of $\zeta$ for the SNM (dotted lines) and PNM (solid lines). The
results for the $\epsilon(\zeta)$ are obtained for $\zeta =0 - 0.06$
by using the parameter sets FSUGZ00 $-$ FSUGZ06.  The differences
$\Delta \epsilon$ at higher densities are very sensitive to the value
of  $\zeta $.}

\caption{\label{fig:snm_pnm} The EOS for the (a) SNM and (b) PNM for the
FSUGZ03 parametrization are compared with the ones calculated within the
DBHF framework  for a realistic potential \cite{Li92}. Similar results
for the NL3 and FSUGold parametrizations are also shown.}

\caption{\label{fig:be} Relative errors   in the total binding energy
$\delta B = (B^{exp}-B^{th})/B^{exp}$ for the newly generated parameter
set FSUGZ03. For the sake of comparison, the values of $\delta B$ obtained
for NL3 and FSUGold parameter sets are also displayed. The  rms errors
in the total binding energies obtained by considering the nuclei used
in  our fits are 1.6, 2.5 and 3.6 MeV for the FSUGZ03, NL3 and FSUGold
parameter sets, respectively.  }

\caption{\label{fig:rch}
Relative errors   in the charge rms radii $\delta r_{ch}$ for the newly
generated parameter set FSUGZ03. Similar results obtained using
NL3 and FSUGold parameter sets are also plotted. The  rms errors in the
charge rms radii  obtained by considering the nuclei used in  our fits
are 0.03, 0.02 and 0.03 fm for the FSUGZ03, NL3 and FSUGold parameter
sets, respectively.  }

\caption{\label{fig:nskin}
Comparison of the results for the neutron-skin thickness $\Delta r = r_{n}
- r{_p}$ obtained using the FSUGZ03 , NL3 and FSUGold parameter sets. The
crosses are  the recent predictions for $\Delta r$ based upon the DBHF
calculations \cite{Alonso03} and isospin diffusion data \cite{Chen05}.}

\caption{\label{m-r} Relation between the neutron star mass and its
radius $R$ for the FSUGZ00, FSUGZ03,FSUGZ06, FSUGold and NL3
parametrizations. The various constraints as indicated by causality,
rotation and $\Delta I/I = 0.014$ are discussed in the text.}

\caption{\label{fig:zeta_mmax} Variation of the limiting mass ${\it
M}_{max}$ of the neutron star with the  OMSC parameter $\zeta$.
For different values of $\zeta$ the remaining parameters of the ERMF model
are adjusted to give best fit to the properties of the finite nuclei.}

\end{figure}
\begin{figure}
\caption{\label{yp_rho} Proton fractions as a function of density for
the FSUGZ00, FSUGZ03, FSUGZ06, FSUGold and NL3  parametrizations. The
solid circles indicate the threshold density for the direct Urca process.}

\caption{\label{fig:zeta_rhouc}
The density $\rho_{\scriptstyle U}$ at which direct Urca process sets
in and the central density $\rho_c$ for the neutron stars with $1.4{\it
M}_\odot$ as a function of OMSC  parameter $\zeta$.  }

\caption{\label{fig:zeta_mu} The minimum value of the neutron star
mass ${\it M}_U$ in which direct Urca process occurs as a function OMSC
parameter $\zeta$.  }

\end{figure}
 
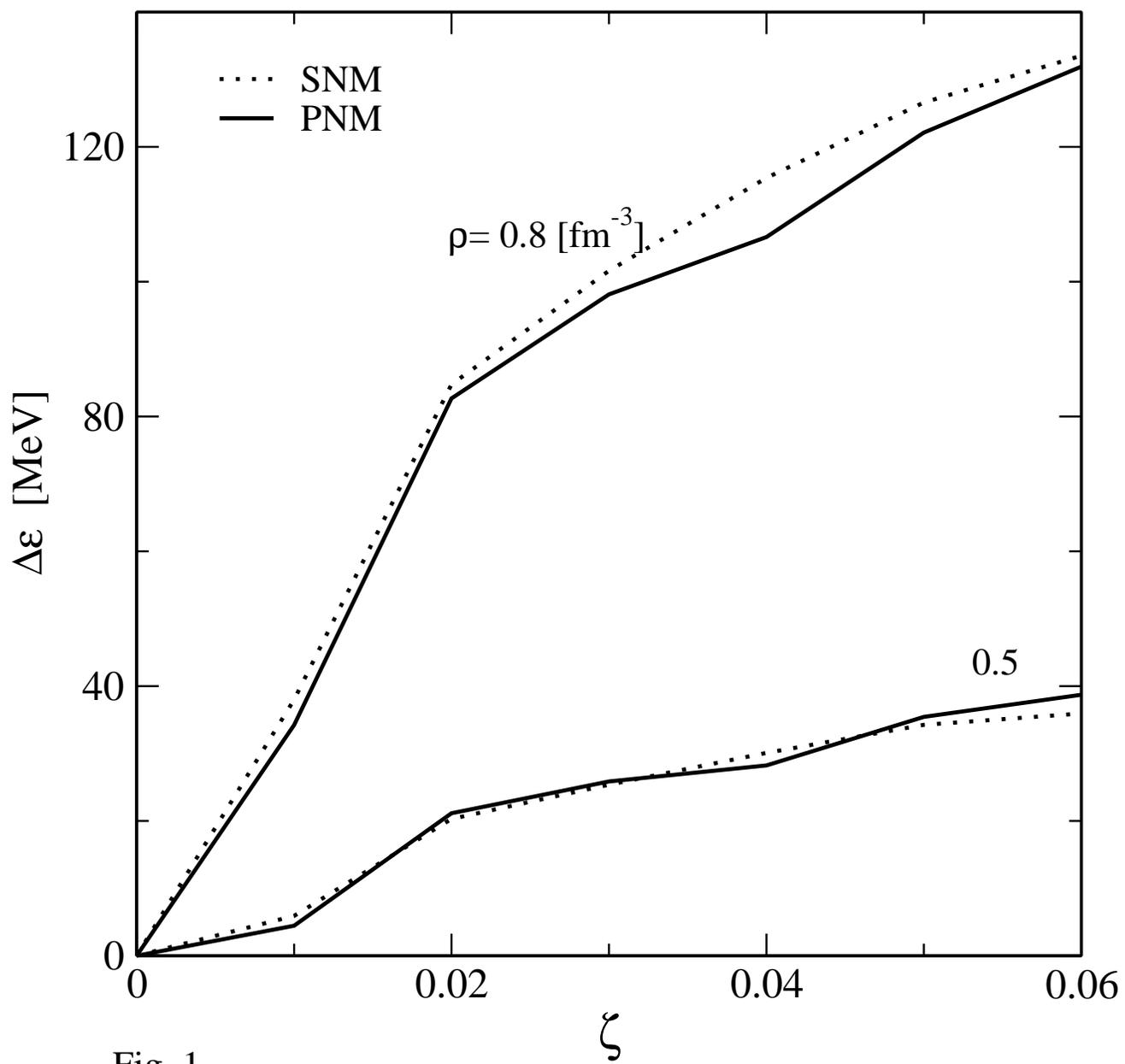
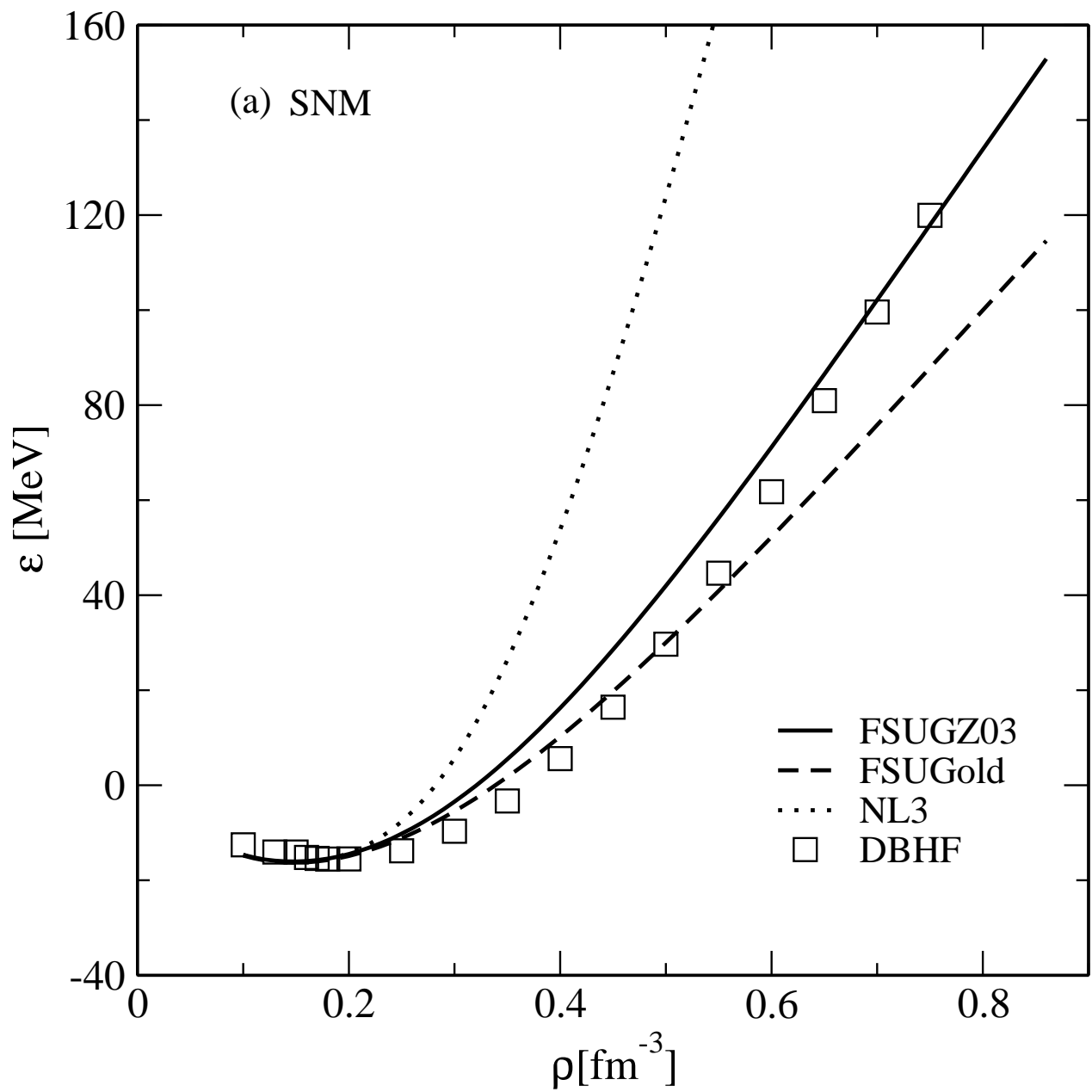
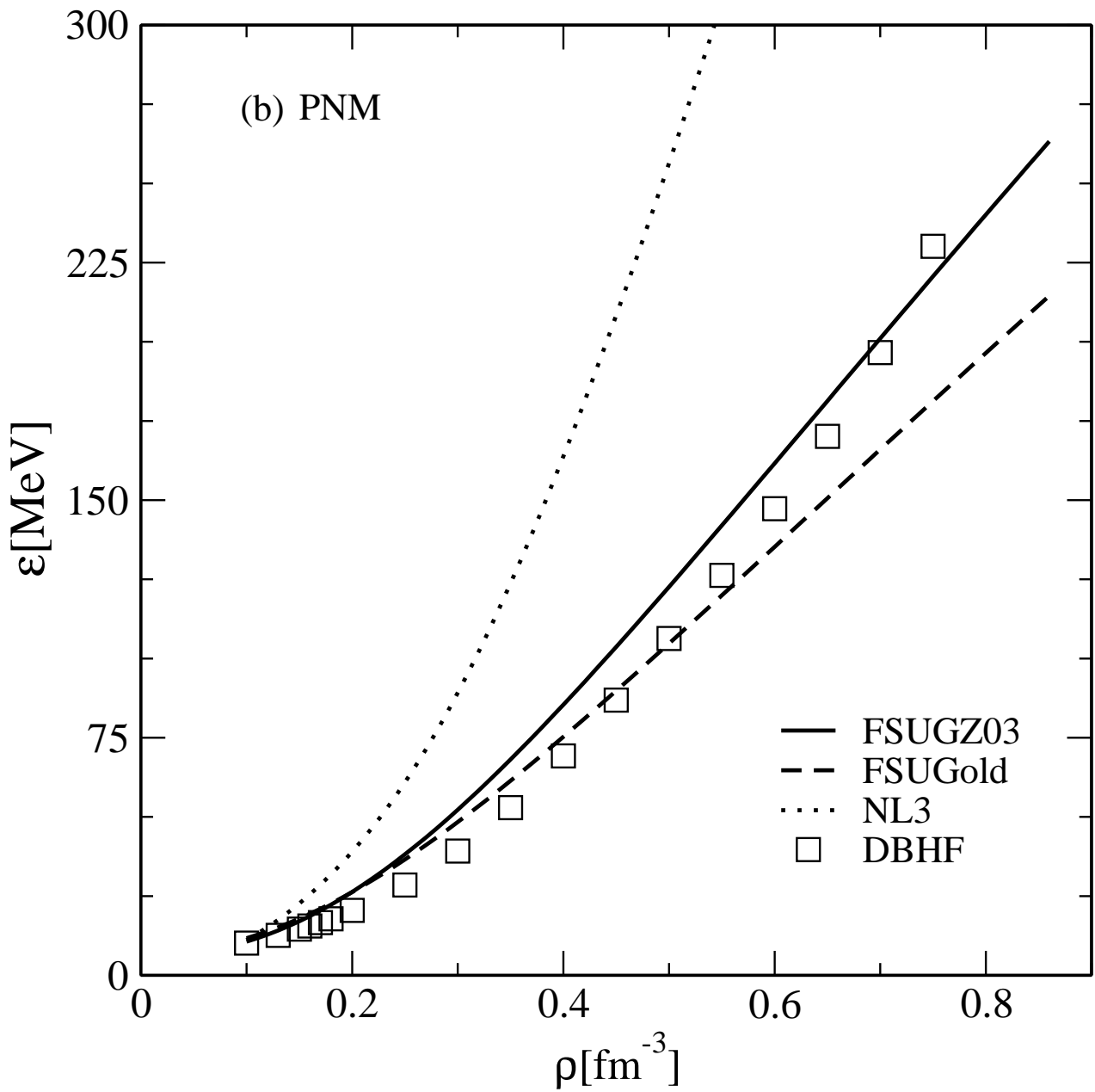
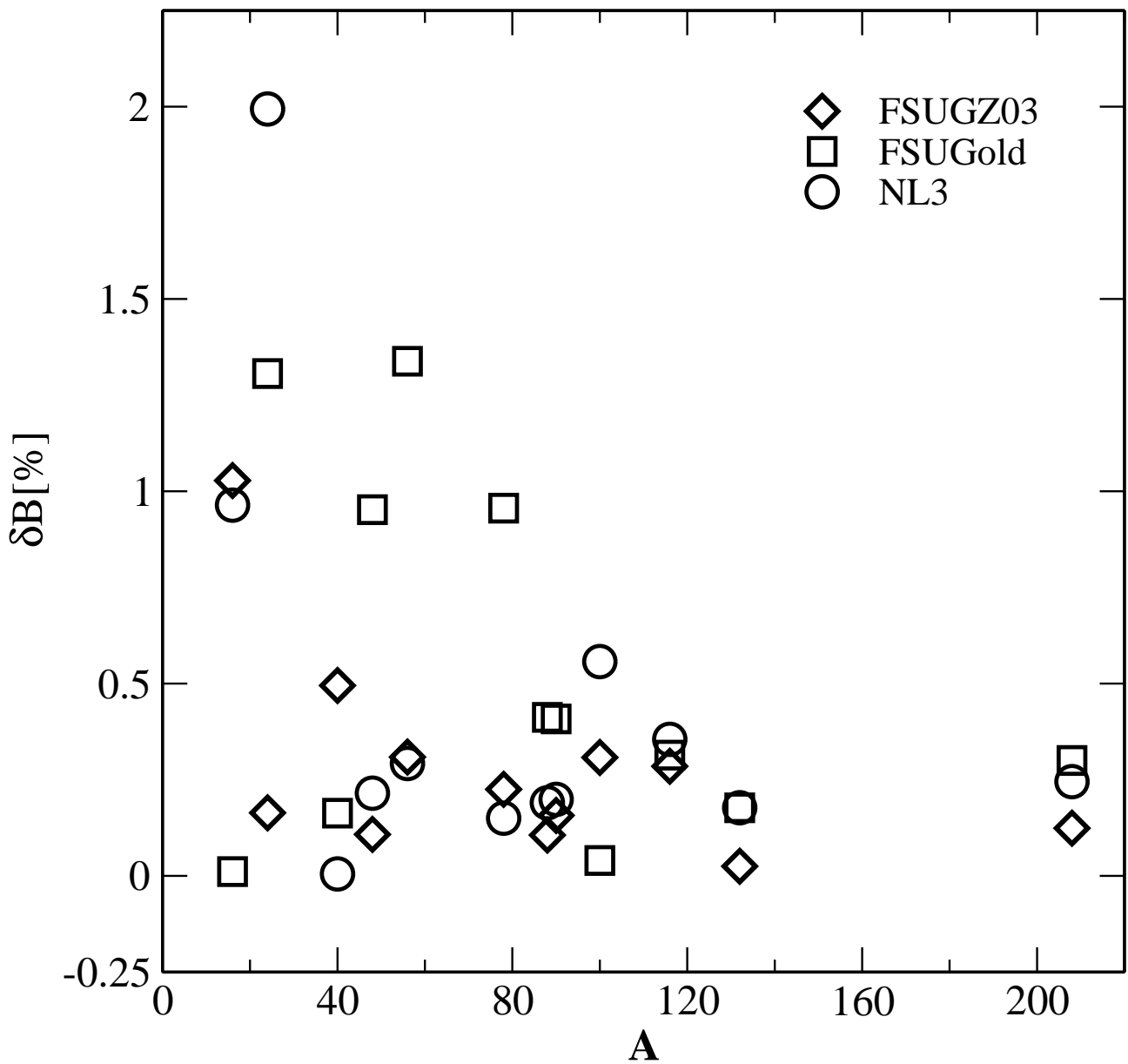
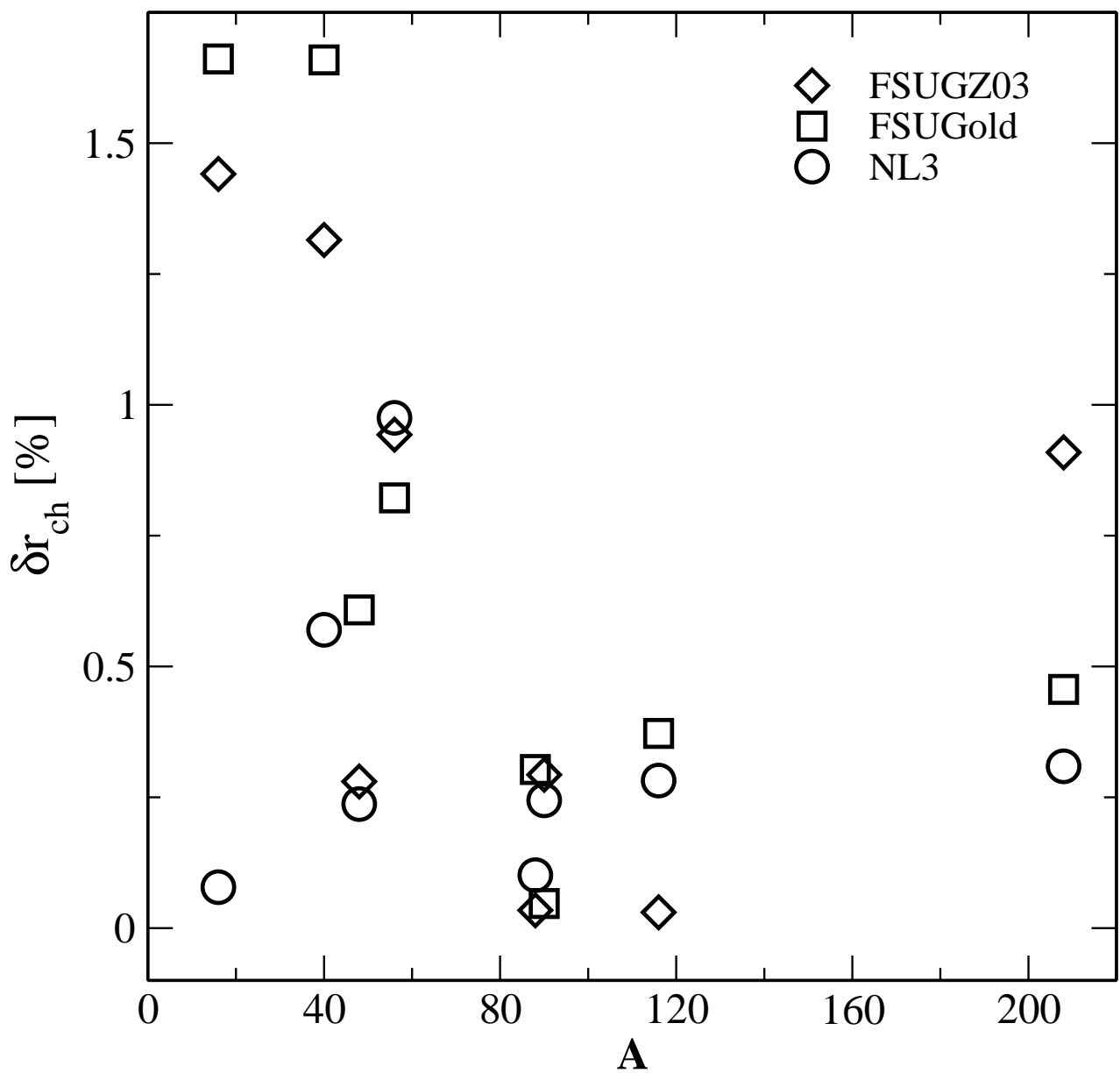
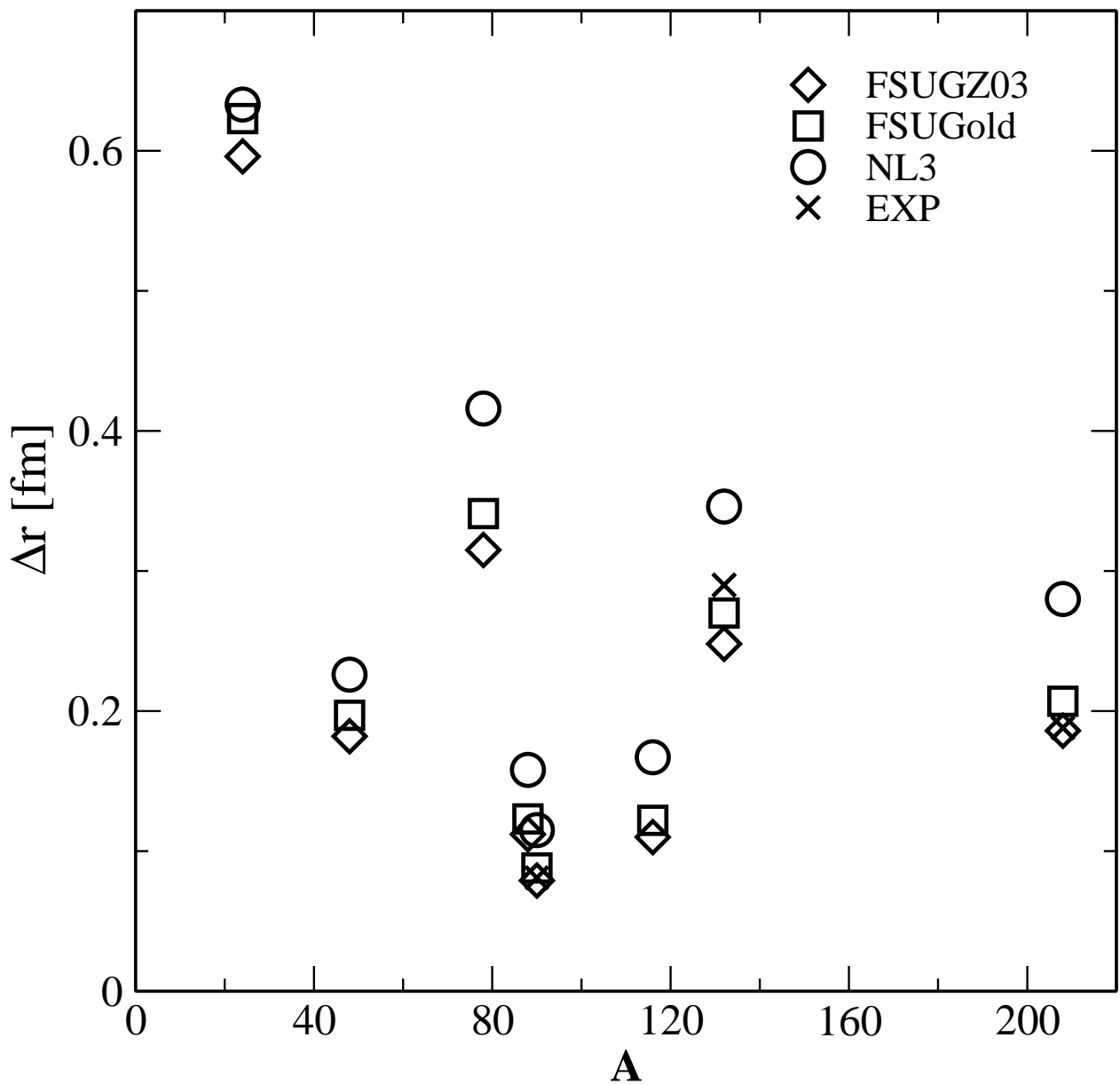
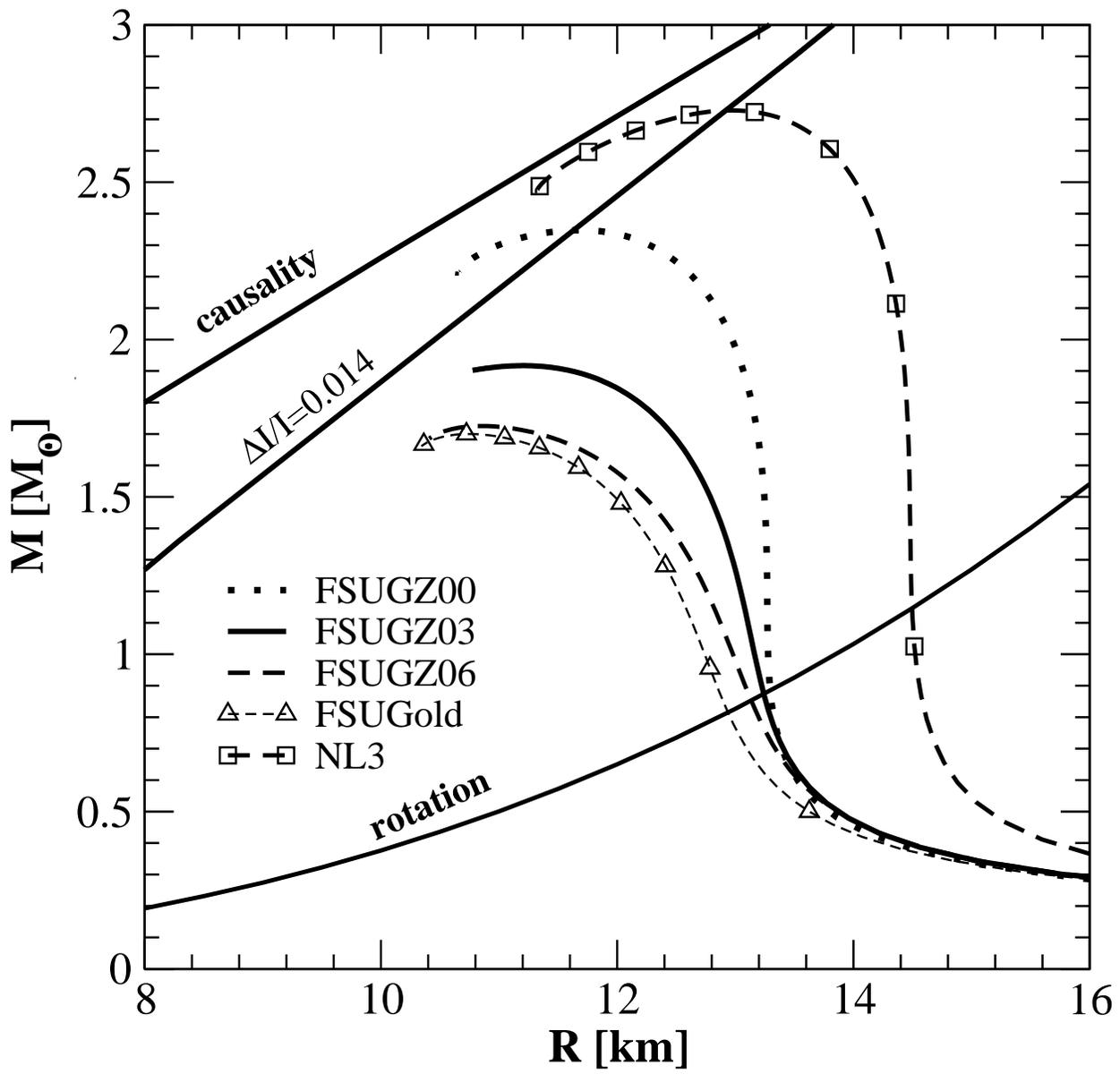
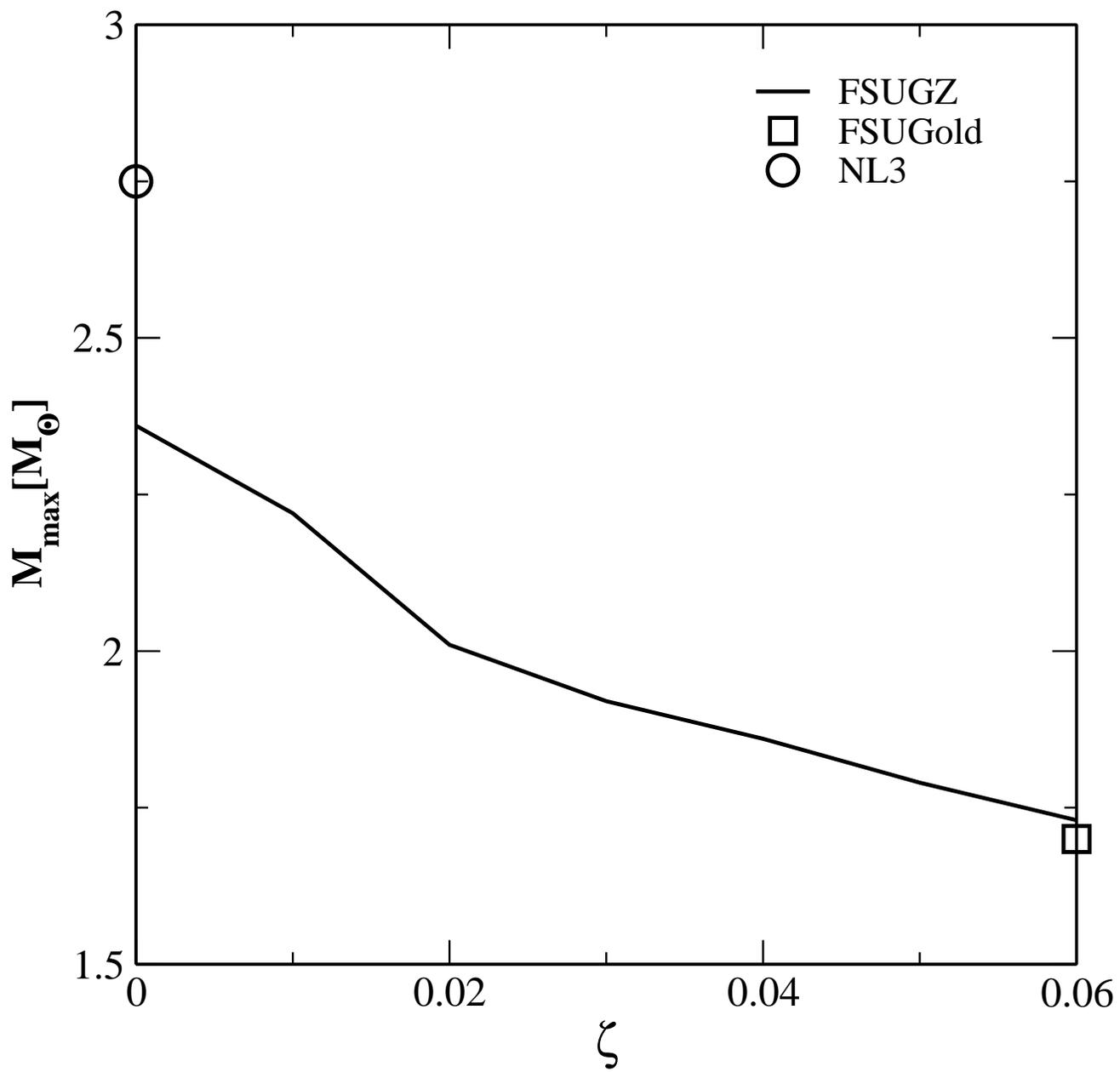
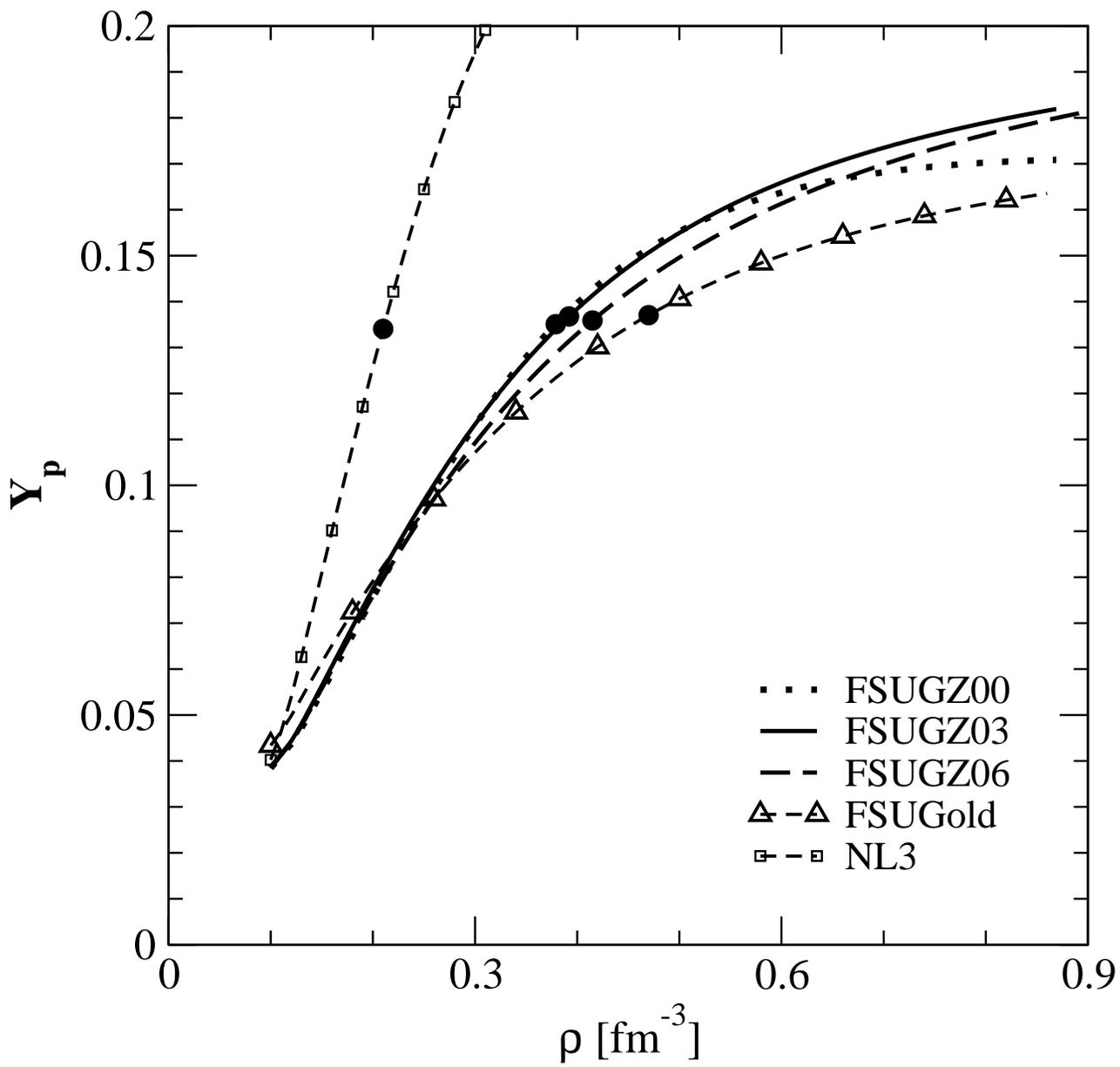
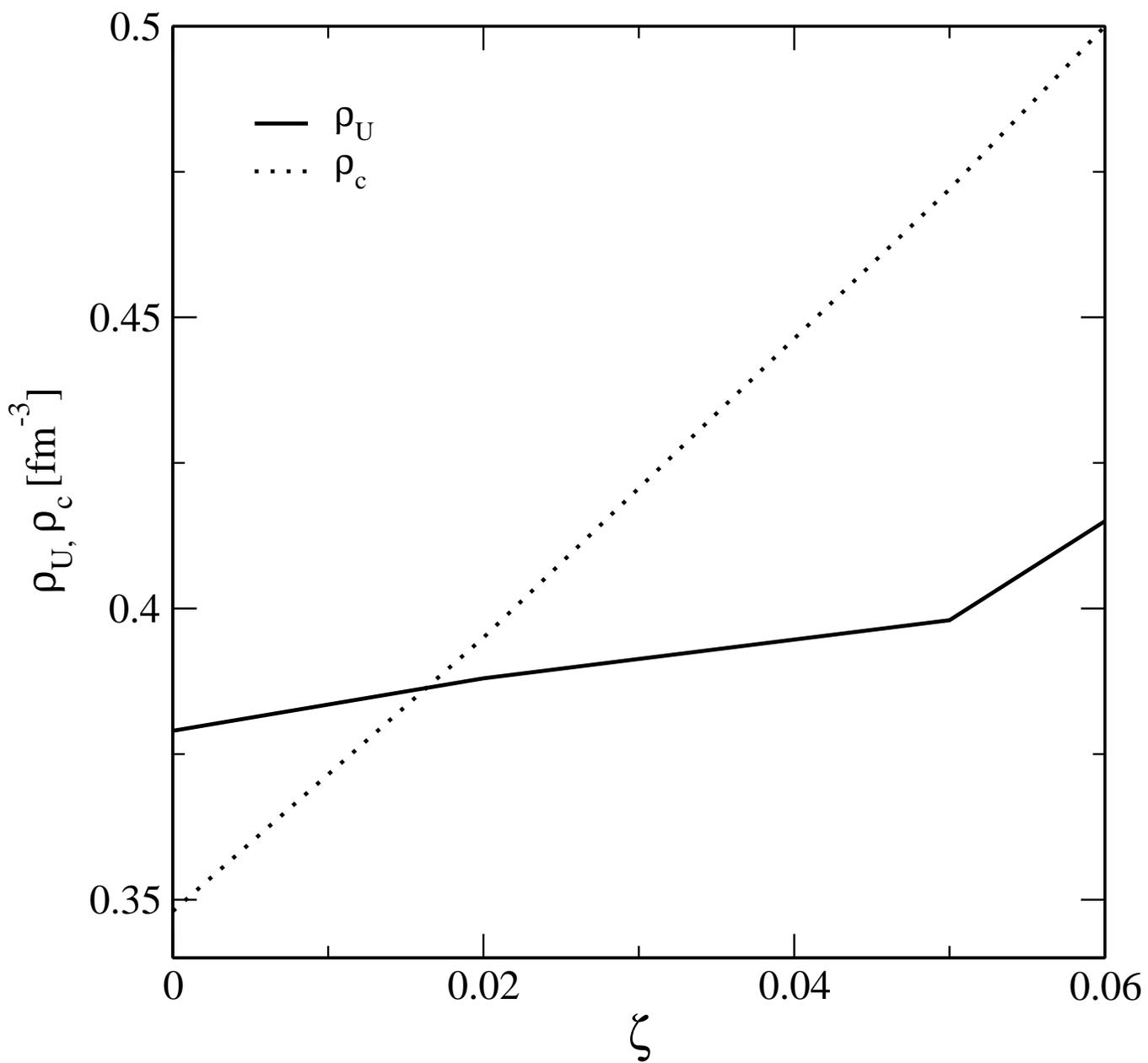
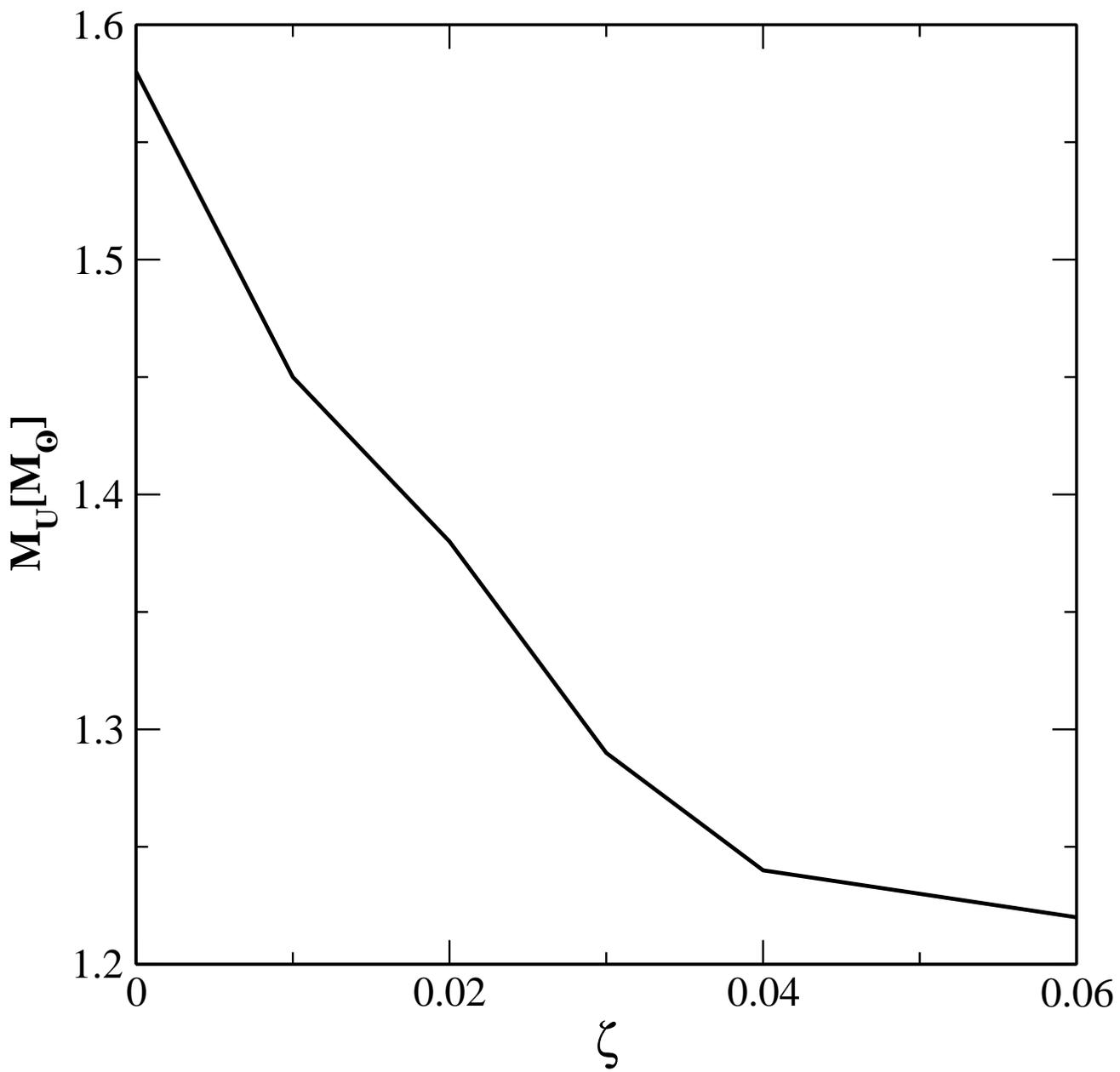
\end{document}